\pdfoutput=1
\documentclass[a4paper,american,floatfix,pdftex,superscriptaddress,twoside,%
aps,pra,
twocolumn,reprint
]{revtex4-2}%
\usepackage{amsmath,amssymb}
\usepackage{lmodern}
\usepackage{graphicx}
\usepackage[T1]{fontenc}
\usepackage[utf8]{inputenc}
\usepackage{microtype}
\usepackage{textcomp}
\usepackage[obeyFinal=True,textsize=footnotesize]{todonotes}
\usepackage{xspace}
\usepackage{hyperref, hypernat}
\usepackage{braket}

\newlength{\figwidth}
\newcommand*{\degree}[1]{\ensuremath{\ifx\\#1\\\else\xspace#1\,\fi^\circ}\xspace}%

\newcommand{\cfeldesy}{\affiliation{Center for Free-Electron Laser Science, Deutsches
      Elektronen-Synchrotron DESY, Notkestrasse 85, 22607 Hamburg, Germany}}%
\newcommand{\uhhcui}{\affiliation{The Hamburg Center for Ultrafast Imaging, Universität Hamburg,
      Luruper Chaussee 149, 22761 Hamburg, Germany}}%
\newcommand{\uhhphys}{\affiliation{Department of Physics, Universität Hamburg, Luruper Chaussee 149,
      22761 Hamburg, Germany}}%

\begin{document}
\title{Controlling rovibrational state populations of polar
	molecules in inhomogeneous electric fields of the Stark deceleration: molecular dynamics and quantum chemistry simulations}%

\author{Emil Zak}
\email{emil@beit.tech}
\altaffiliation{Current affiliation: \href{https://beit.tech/}{BEIT Inc.}, \\Mogilska 43, 31-545 Krak\'ow, Poland}
\cfeldesy

\author{Jochen Küpper}
\cfeldesy
\uhhcui
\uhhphys

\author{Andrey Yachmenev}
\email{andrey.yachmenev@robochimps.com}
\cfeldesy

\date{\today}%

\begin{abstract}\noindent%
 We propose a modified Stark-chirped rapid adiabatic passage technique for a robust rovibrational population transfer in the gas 
 phase molecules in the presence of certain inhomogeneous electric fields. As an example application, the new state switching 
scheme is shown to greatly enhance the process of slowing polar ammonia molecules in the Stark decelerator. High-level 
 quantum mechanical simulations show that a virtually complete population inversion between a selected pair  of 
 weak-field and  strong-field seeking states of NH$_3$ can be achieved. Strong dc electric fields within the Stark decelerator are 
 used as part of the rovibrational population transfer protocol. 
 Classical-dynamics simulations for ammonia demonstrate notable improvements
 in the longitudinal phase space acceptance of the Stark decelerator upgraded with the state switching and an increased
 deceleration efficiency with respect to the standard Stark deceleration technique.
\end{abstract}
\maketitle%

\section{Introduction}

Robust control of rotational-vibrational (rovibrational) state populations in the gas phase molecules brings a number 
of applications, ranging from quantum state-selection \cite{Perreault:PRL124:163202,Oberst:PRA78:033409} for  
fundamental physics studies, enhanced 
lasing~\cite{Li:PRL125:053201} or implementation of quantum gates~\cite{Mller:PRL100:170504}, to mention few. 

A coherent population transfer between a pair of rovibrational states can be achieved with near unit efficiency by the 
Stark-chirped rapid adiabatic passage (SCRAP)
technique~\cite{Rickes:JCP113:534, Rangelov:PRA72:053403, Oberst:PRA78:033409}. Originally, SCRAP
implements two time-delayed Gaussian pulses that partly overlap in time: A pump pulse with a fixed
frequency is tuned slightly away from the resonance with the transition of interest. A relatively intense far-off-resonance Stark 
pulse is used to chirp the
transition frequency  through the resonance.
Here we propose a modification of the 
SCRAP method, in which the role of the Stark pulse is taken by the dc electric fields present in the 
system of interest.  We demonstrate how the population switching can 
greatly 
enhance the process of slowing polar molecules, on the example of the Stark decelerator~\cite{Bethlem:PRL83:1558, 
Meerakker:NatPhys4:595, 
	Meek:PRL100:153003,
	Bell:MP107:99, Hogan:PRL103:123001, Meerakker:CR112:4828}. Yet the scheme presented here is 
designed to have applications ranging beyond the Stark deceleration technique. 

\begin{figure}[b]
   \includegraphics[width=\linewidth]{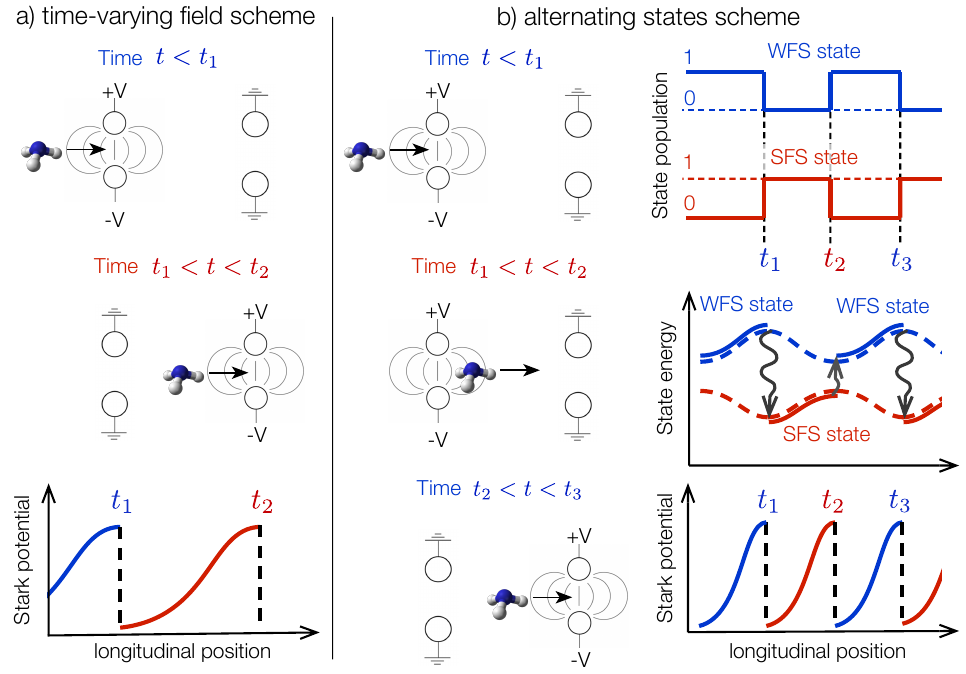}
   \caption{Stark potential energy of polar molecules as a function of the longitudinal position
      along the molecular beam axis through the electrode pairs. An effective moving potential wall
      is created in front of molecular beam by (a) repeated switching of the voltages between field
      configurations at times $t_1$ and $t_2$ or (b) repeated altering of the molecular quantum
      state between selected SFS and WFS states at times $t_1$, $t_2$ and $t_3$ and corresponding
      switching of voltages between field configurations at times $t_1,t_2,t_3,\ldots$}
   \label{fig:1}
\end{figure}

Stark deceleration~\cite{Bethlem:PRL83:1558, Meerakker:NatPhys4:595, Meek:PRL100:153003,
   Bell:MP107:99, Hogan:PRL103:123001, Meerakker:CR112:4828} is a popular modern approach to
cold-molecules experiments~\cite{Veldhoven:EPJD31:337, Krems:IRPC24:99, Gilijamse:Science313:1617,
   Hudson:PRL96:143004, Schnell:FD150:33, Stuhl:ARPC65:501, Wall:NJP17:025001,
   Vogels:Science350:787}, where it is used for accelerating, decelerating, and guiding packets of
state-selected small polar molecules. The operation principle of Stark deceleration is illustrated
in \autoref{fig:1}b. The inhomogeneous electric field, produced by voltage differences between
pairs of opposing electrodes, creates an effective Stark potential for polar molecules. For
molecules in a weak-field seeking (WFS) internal state, the potential energy increases along the
molecular beam axis. It reaches the maximum at the position of the maximum field strength, i.e.,
between the electrodes. Molecules moving in the direction of electrodes climb the potential hill
and, therefore, loose kinetic energy. At a moment before molecules reach the position of maximum
electric field, the voltages are quickly switched. This prevents molecules from regaining their
kinetic energy on the downward slope of the Stark potential. Instead, they have to climb the next
hill. This process is repeated by letting the molecules pass through multiple electric field stages.

Conversely, molecules in a strong-field seeking (SFS) state exhibit a Stark potential that has a
form of a well with its minimum energy at the position of the maximum electric field. Therefore,
molecules will gain the kinetic energy when moving in the direction of increasing electric field
(toward electrodes) and loose it when going away. In principle, such molecules can be also focused
and decelerated by appropriately switched fields~\cite{Auerbach:JCP45:2160, Bethlem:PRL88:133003,
   Tarbutt:PRL92:173002, Bethlem:JPB39:R263, Wohlfart:PRA77:031404R, Wohlfart:PRA78:033421,
   Filsinger:PRL100:133003, Chang:IRPC34:557}.

One may think of switching the population from one quantum state to another at the points of minimum
and maximum of the electric field, such that the effective Stark potential is always upwards-sloped
in the direction of the molecular motion. In this way, molecules will continually loose their
kinetic energy on both the upward and downward slopes of the electric field, which is illustrated in
\autoref{fig:1}b. A seemingly similar idea to Stark deceleration has been theoretically proposed in~\cite{Hudson:PRA79:061407}, 
yet there are a 
number of key differences discussed in section~\ref{sec:population} which make the present technique more robust and more 
practical. Sisyphus opto-electrical
cooling method~\cite{Zeppenfeld:Nature491:570, Prehn:PRL116:063005} is also based on a similar idea to
produce ultra-cold molecules. Very recently, a concept of state switching with adiabatic-passage techniques has been 
utilized~\cite{Mukherjee:APL116:251103} to control the excition qubit dynamics in quantum dots. 

In the Stark decelerator, the field switching sequence is calculated with the help of a hypothetical
synchronous molecule moving along the molecular beam axis. The fields are switched every time the
synchronous molecule reaches a certain position with respect to the electrodes, expressed in terms
of a phase angle $\phi_0=0\ldots\degree{90}$. The limiting values 0 and $\degree{90}$ correspond to
the points of minimum and maximum of electric field along the beam axis. The phase angle is always
chosen slightly away from the position of the field maximum ($\degree{\phi_0\approx80}$) in order to
increase the phase space, i.e., velocity and position, acceptance of molecules along the molecular
beam axis. The exact value of the phase angle is calculated depending on the initial position and
velocity distribution of molecules and the desired final velocity. In the transverse directions,
molecules in WFS states are focused towards the molecular beam axis, resulting in time-dependent
transverse oscillatory trajectories of molecules around the synchronous molecule.

In the alternating states (AS) Stark deceleration method, see \autoref{fig:1}, the electric field
configuration is switched at $\phi_0\approx0$, while the population is transferred back and forth
between the WFS and SFS states at $\phi_0\approx0$ and \degree{90}. The transverse focusing
mechanism is similar to that in the alternating gradient decelerator~\cite{Bethlem:JPB39:R263},
except the fact that molecules are focused toward the beam center in WFS state and defocused away in
the SFS state moving along the same transverse direction and within one electric field stage.

Besides Stark deceleration, several other techniques for producing slow molecular beams have been
advanced in the recent years, such as buffer-gas cooling~\cite{Hutzler:CR112:4803,
   Singh:PRA97:032704}, direct laser cooling~\cite{Barry:PRL108:103002, Truppe:NJP19:022001},
Sisyphus~\cite{Zeppenfeld:Nature491:570, Comparat:PRA89:043410, Prehn:PRL116:063005},
sympathetic~\cite{Lim:PRA92:053419}, or evaporative~\cite{Stuhl:Nature492:396} cooling, and
photoassociation of ultracold atoms~\cite{Ni:Science322:231, Perez:PRL115:073201}. Many of these
techniques can benefit from using the Stark decelerator as a preceding
stage~\cite{Cheng:PRL117:253201} for producing high intensities of cold beams in selected
rovibrational states from a wide range of polar molecular species.

Here, we explore the idea of altering the internal state of the whole ensemble of molecules in the molecular beam in the presence 
of inhomogeneous electric fields. As an example we show how this method help to improve the
efficiency and the phase-space characteristics of the Stark deceleration method. We propose and
theoretically validate a robust method for population inversion between a pair of SFS and WFS states
of a molecule, based on the rapid-adiabatic passage with a Stark-shift chirp produced by the
controlled time-modulation of the electric field in Stark decelerator. Applied to a beam of ammonia
($^{14}$NH$_3$) molecules, the proposed method shows 2-3 times larger longitudinal velocity
acceptance than the conventional decelerator resulting in higher final densities of cold molecules.
Furthermore, since molecules are slowed down on both upward and downward slopes of the electric
field, the number of mechanical deceleration stages, required to reach a desired final velocity, can
be reduced by a factor of two. This aids the deceleration of heavy or less polar molecules at lower field strengths. The
present method could be implemented using one of the existing Stark decelerators by mounting an
additional laser source and fast semiconductor voltage switches.

\section{Population inversion in the Stark decelerator}
\label{sec:population}
A coherent population transfer between a pair of WFS and SFS rovibrational states can be produced
with a very high efficiency by the Stark-chirped rapid adiabatic passage (SCRAP)
technique~\cite{Rickes:JCP113:534, Rangelov:PRA72:053403, Oberst:PRA78:033409}. The standard SCRAP
implements two time-delayed Gaussian pulses that partly overlap in time: A pump pulse with a fixed
frequency $\omega_p$ is tuned slightly away from the resonance with the WFS-to-SFS transition
$\Delta{}E_\text{res}$. A relatively intense far-off-resonance Stark pulse is used to chirp the
transition frequency $\Delta{}E_\text{res}+\Delta_\text{S}(t)$ through the resonance with
$\hbar\omega_p$ by inducing a dynamic Stark shift $\Delta_\text{S}(t)$. The SCRAP population
transfer is robust to fluctuations in the intensities of the pulses, as long as other parameters --
such as the pump frequency, and the delay time -- are carefully controlled.

It is therefore essential that there are no further strong detunings present in the system, e.g.,
inhomogeneous broadenings of transition frequency or external electric fields. Such perturbations
can significantly lower the percentage of the population transfer, which represents a major obstacle
for employing SCRAP within the Stark decelerator. Although the highly spatially inhomogeneous
electric field aids the deceleration, it induces an additional static detuning of the transition
frequency, i.e., $\Delta{}E_\text{res}+\Delta_\text{dc}(\vec{r})$, which depends on the position of
the molecule $\vec{r}$ in the decelerator. As a result, SCRAP will produce a complete population
inversion only for a fraction of the total amount of molecules, which have an appropriate dc Stark
shift, i.e., only for molecules located within certain confined regions in the decelerator. This
would, for example, demand two different sets of the Stark and pump pulses to be employed for the
population inversion at the areas of minimum and maximum of the electric field. However, the
principal problem is that the variation in the dc field detuning over a typical spatial-acceptance
region of the decelerator is much larger than the detuning normally produced by a Stark pulse,
i.e., $|\Delta_\text{dc}(\vec{r'})-\Delta_\text{dc}(\vec{r})|\gg\Delta_\text{S}(t)$. Nonetheless,
increasing the Stark pulse intensity, i.e., increasing $\Delta_\text{S}(t)$, would make the
population transfer more sensitive to the intensity fluctuations in the pump pulse. One can try to
make $\Delta_\text{dc}(\vec{r})=0$, i.e., switch off the dc electric field for a short period of
time, i.e., a fraction of a microsecond, and carry out the SCRAP population inversion. This, however,
may lead to the so-called Majorana transitions into the non-polarizable molecular states, e.g.,
states with $m=0$, and correspondingly to a significant loss of molecular density. Another option
would be to create a dc electric field that is uniform in the longitudinal and transverse
directions, i.e., $\Delta_\text{dc}(\vec{r})=\text{const.}$, which is, however, technically very
challenging; it could possibly be implemented in a Stark decelerator with large-flat-area
electrodes~\cite{Hudson:PRA79:061407}. 

Here, we develop an alternative scheme to circumvent this problem by using a temporally controlled
rise of the dc electric field in decelerator to produce the rapid-adiabatic passage transfer. The
second Stark pulse of the original SCRAP is discarded and the operation scheme can be described as
follows: when molecules reach a position where their population has to be transferred between SFS
and WFS states, see \autoref{fig:1}, the voltages $V(t)$ are first dropped to $V_\text{min}$ as in
the conventional Stark decelerator. Then the pump field ($\omega_p$) is turned on and the voltages
$V(t)$ are ramped up to $V_\text{max}$ in a time of about 0.1--0.2~$\mu$s
($V_\text{min}\leq{}V(t)\leq{}V_\text{max}$). The applied voltages generate the time- and
position-dependent dc electric field, which in turn produces the spatially inhomogeneous Stark shift
of the transition frequency $\Delta_\text{dc}(\vec{r},t)$ in molecules. The minimal and the maximal
voltages are chosen such that by varying $V(t)$ over a period of time $t=0\ldots{}t_\text{max}$, the
condition
\begin{eqnarray}
  \Delta E_\text{res}+\Delta_\text{dc}(\vec{r},t_\text{res}) = \hbar\omega_p
\end{eqnarray}
is satisfied at some resonance time $0<t_\text{res}<t_\text{max}$ for every position $\vec{r}$
within a large acceptance-volume inside the decelerator. The sufficiently intense pump field,
applied during the resonance crossing, produces a complete adiabatic passage of population from
initial to final states~\cite{Rangelov:PRA72:053403}. The adiabatic evolution of the population is
ensured by a smooth temporal rise of the voltages, a long interaction time, and a strong electric
dipole coupling between the interacting states. When the voltages reach the operating value
$V_\text{max}$, the pump field is turned off and molecules continue into the next deceleration
stage. In ref.~\cite{Hudson:PRA79:061407} a seemingly similar idea of WFS/SFS population 
switching  has been proposed, however there are some key differences. Firstly, in~\cite{Hudson:PRA79:061407} two laser pulses: 
Stark pulse and pump pulse are required. Secondly, ref.~\cite{Hudson:PRA79:061407} uses stimulated Raman adiabatic-passage 
method to switch between strong-field
seeking and weak-field seeking states, whereas we use Stark-Chirped rapid-adiabatic passage (SCRAP) method. A disadvantage 
of STIRAP is that its efficiency in transferring populations between rovibrational states is sensitive to
inhomogeneous broadening of respective energy levels~\cite{Oberst:PRA78:033409}. To the contrary, our Stark-chirped
rapid-adiabatic-passage method \emph{utilizes} the Stark-shifts to rovibrational energy
levels as a means of sweeping through resonance with the pump-field frequency. In this way, there is a reasonably large margin 
for energy dispersion of the initial/final state.

We simulate the SCRAP population transfer in the molecule by solving the corresponding
time-dependent Schrödinger equation and calculating the time-dependent probabilities for finding the
molecule in the SFS or WFS state. Specifically, the objective is to optimize the time-dependent
voltage function $V(t)$ that would maximize the efficiency of the population transfer between
selected rovibrational states. For these simulations we choose the prototypical ammonia
($^{14}\text{NH}_3$) molecule. The lowest-energy state of para-ammonia is the rotational state
$|J,k,m\rangle=|1,1,1\rangle$ of the vibronic ground state. This state is split by
$\Delta{E}_{a-s}=0.794$~cm$^{-1}$ due to the inversion tunneling into the symmetric
$E_s=16.173$~cm$^{-1}$ and the antisymmetric $E_a=16.967$~cm$^{-1}$ component. These components have
opposite parity and repel each other in the presence of an external electric field, producing a pair
of closely lying SFS and WFS states. To model the dc electric field produced by the Stark
decelerator we picked a conventional electrode configuration. Each stage of the decelerator is
formed by two parallel cylindrical electrodes of diameter 3~mm, with their axes parallel to either
$x$ or $y$, and their surfaces separated by 2~mm. The distance between the centers of two stages is
6 mm along the molecular beam axis $z$, and successive stages are rotated through $90^{\circ}$ about
the $z$ axis. When voltage is applied to electrodes along the $x$ axis, the electrodes along the $y$
axis are grounded and vice versa, such that the electric field vector is always contained in the
$xz$ or $yz$ planes. Since the electric field is periodic and symmetric, for simulations we choose a
unit cell $z=[z_\text{min},z_\text{max}]$ between the centers of two neighbouring stages, where the
electrodes at $z_\text{min}$ are grounded and the voltage is applied to the electrodes at
$z_\text{max}$. The resulting dc electric field is computed on a dense grid using a finite elements
approach~\cite{Comsol:Multiphysics}. The maximal value of the applied voltage was set to
$\left|V_\text{max}\right|=10$~kV, which creates an electric field of 1.9~kV/cm, 23.2~kV/cm, and
91.1~kV/cm at $z_\text{min}$, $(z_\text{min}+z_\text{max})/2$, and $z_\text{max}$, respectively,
along the molecular beam axis through the electrode pairs.

To simplify the analysis of the time-dependent probabilities in terms of the SFS and WFS state
populations, the minimal voltage was set to zero, $V_\text{min}=0$, at the end of SCRAP simulation.
In the actual Stark deceleration experiment, however, the electric field is never let below 300~V/cm
to avoid Majorana transitions into the non-polarized $m=0$ states~\cite{Bethlem:PRA65:053416}.

\begin{figure}
   \includegraphics[width=\linewidth]{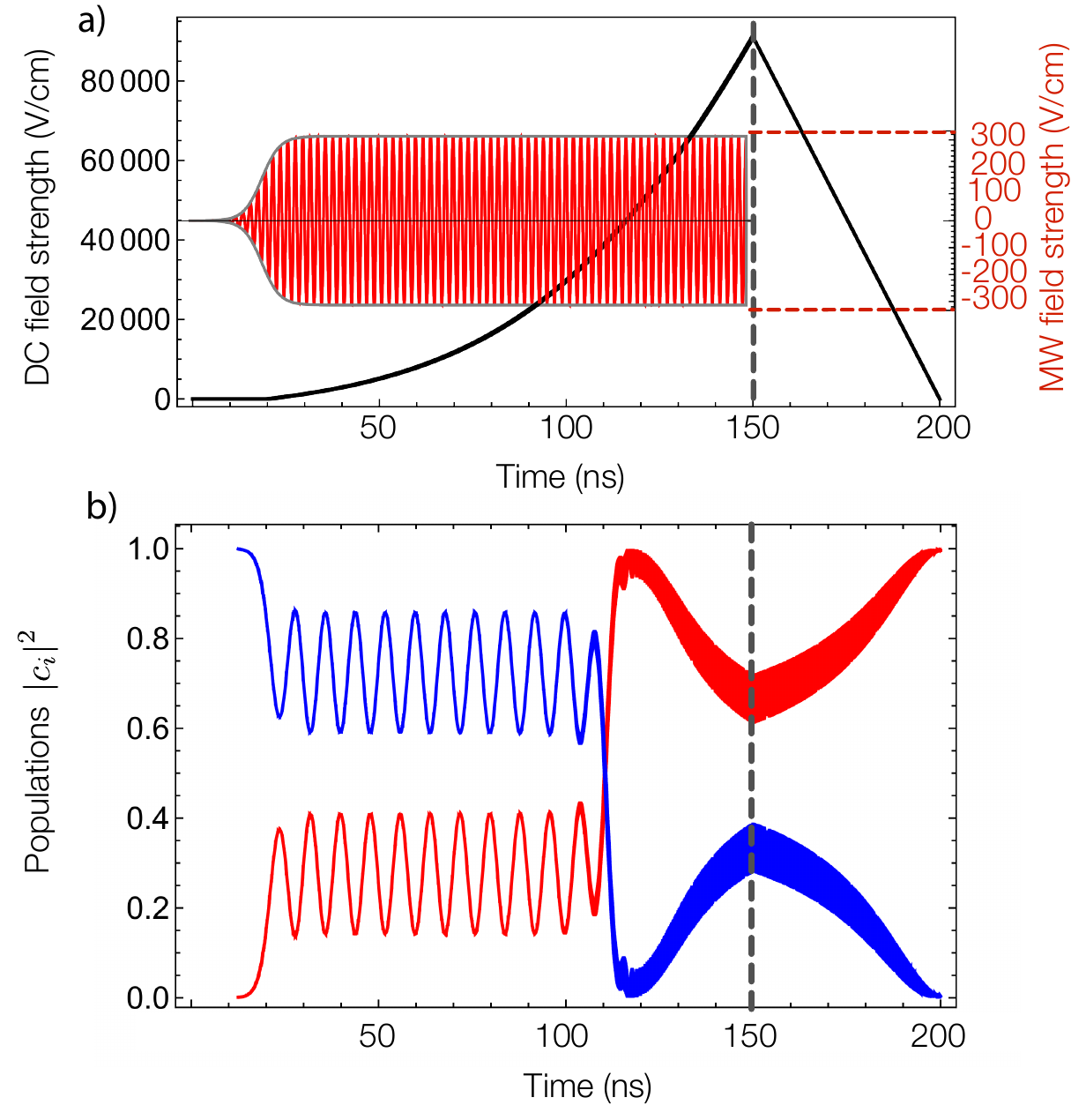}
   \caption{Top: Temporal evolution of the dc (black) and the microwave-pump (red) electric fields.
      The dc field corresponds to the position $z=z_\text{max}$ and the voltage rise function
      parameter $a=0.030$, see~\eqref{eq:1}. Bottom: Temporal evolution of the populations of the
      symmetric (blue) and antisymmetric (red) $\ket{J,k,m}=\ket{1,1,1}$ states of ammonia.}
   \label{fig:2}
\end{figure}

The voltage raise over a time of $t_\text{max}$ is modelled by the function
\begin{equation}
   {V(t) = N^{-1}\ln\left(\frac{1+e^{-a(t_\text{max}-t)}}{1+e^{-at_\text{max}}}\right)(V_\text{max}-V_\text{min})
      + V_\text{min}}
  \label{eq:1}
\end{equation}
where the value of the parameter $a$ is to be optimized and $N=\ln(2/(1+e^{-at_\text{max}}))$ is the
normalization factor. For the present simulations we choose $t_\text{max}=130$~ns. The pump field
\begin{equation}
   E_p(t) =
   \begin{cases}
      \frac{1}{1+e^{-0.4t+8}}E_0\cos\omega_pt & \text{for}~t\leq t_\text{max}, \\
      0 & \text{otherwise},
   \end{cases}
   \label{eq:2}
\end{equation}
is characterized by the frequency $\hbar\omega_p=0.797$~cm$^{-1}$, linear polarization along the
main direction of the electric field, i.e., $x$ or $y$ for different stages, the peak intensity of
$E_0=300$~V/cm, and the duration of $t_\text{max}$. To ensure the adiabatic evolution of the
time-dependent probabilities, the pump pulse time profile is modelled by a quasi-step function with
a smooth sigmoid-shape rise extended over 20~ns. The time evolution of the dc field, generated by
the voltage function in \eqref{eq:1}, together with the pump field $E_p(t)$, are plotted in the top
panel of \autoref{fig:2}, where the voltage rise is delayed by the $E_p(t)$ rising time of 20~ns.

With the use of a full-dimensional spectroscopically refined potential energy surface for
ammonia~\cite{Yurchenko:MNRAS413:1828}, we first variationally obtained the field-free rovibrational
energies and wavefunctions for the rotational quantum number $J=0\ldots10$, including all
vibrational excitations with band centers below 4000~$cm^{-1}$. To model the interaction of the
molecules with the external electric fields, we used a full-dimensional \emph{ab initio}
dipole-moment surface for ammonia~\cite{Yurchenko:JPCA113:11845} and computed all corresponding
rovibrational matrix elements of the laboratory-fixed dipole-moment operator using
TROVE~\cite{Yurchenko:JMS245:126, Yachmenev:JCP143:014105, Yurchenko:JCTC13:4368} and
RichMol~\cite{Owens:JCP148:124102}. The time-dependent wavepacket exposed by interactions with the
electric fields was modelled as a linear combination of the field-free rovibrational states with
time-dependent coefficients. The latter were obtained from numerical solution of the time-dependent
Schrödinger equation (TDSE)~\cite{Owens:JCP148:124102}. The initial wavepacket was given by either
symmetric or antisymmetric $\ket{1,1,1}$ stationary (field-free) state of ammonia.

The populations of different stationary states in the wavepacket are given by the squares of the
respective time-dependent coefficients (probabilities). For example, for the position
$x,y,z=0,0,z_\text{max}$ and the voltage rise parameter $a=0.030~\text{ns}^{-1}$ the results of
numerical simulations are presented in the bottom panel of \autoref{fig:2}. The time evolution of
the populations of symmetric and antisymmetric $\left|1,1,1\right>$ states is plotted.
\begin{figure}
   \includegraphics[width=\linewidth]{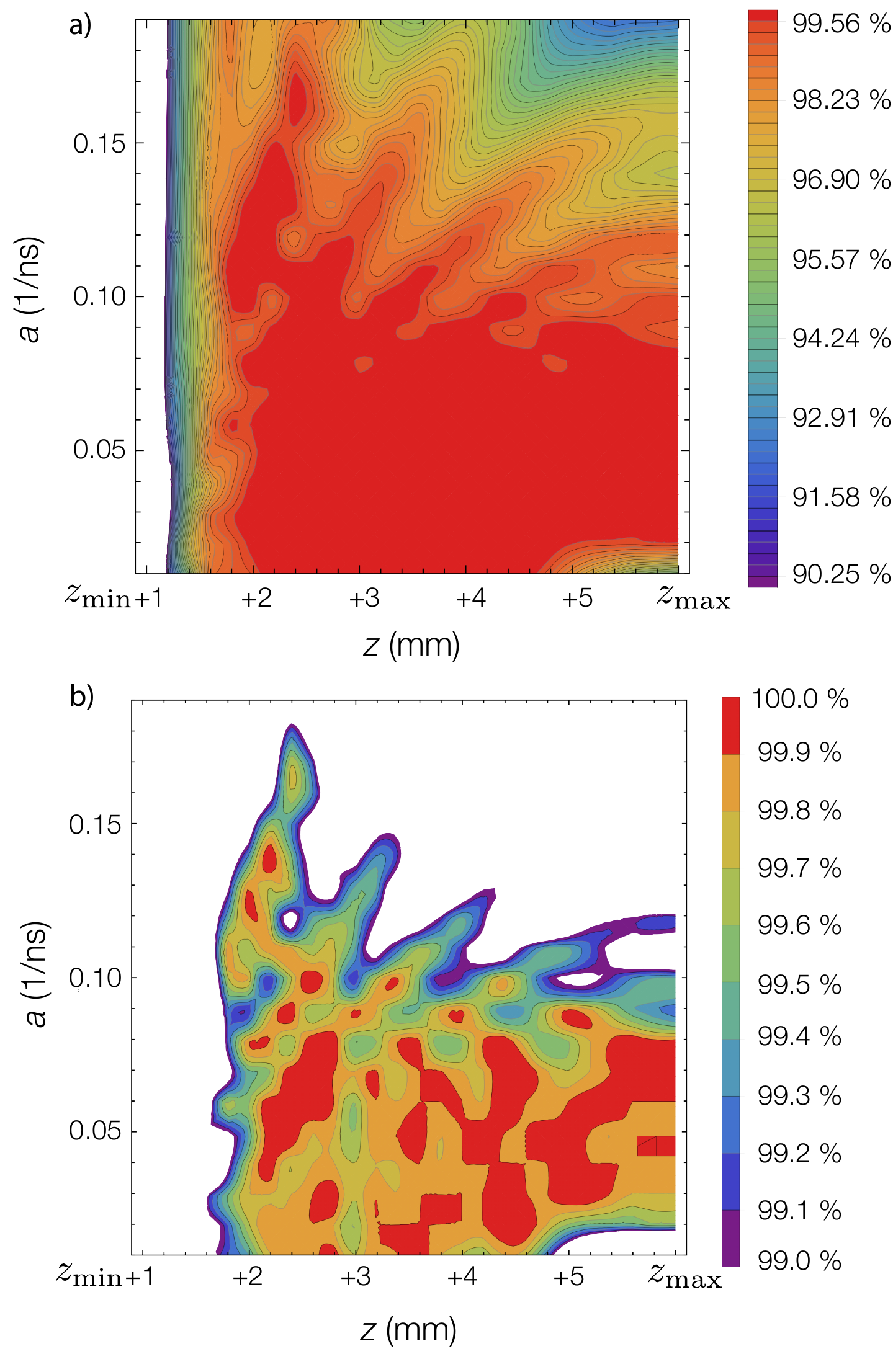}
   \caption{Contour plots of the target state population (in \%) as a function of the voltage time
      rise parameter $a$, as defined in~\eqref{eq:1}, and the longitudinal position $z$ along the
      decelerator axis. Top and bottom plots correspond to population ranges 90-100~\% and
      99-100~\%, respectively.}
   \label{fig:3}
\end{figure}
The corresponding field configuration is sketched in the upper panel of \autoref{fig:2}. At the
terminal time $t_\text{max}=150$~ns the dc electric field reaches its maximum and the SCRAP is
finished. However, since the dc electric field mixes both the symmetric and the antisymmetric state
components, it is not possible to exactly quantify the result of the population inversion effect. To
work out the populations of the stationary states, we continue to propagate the wavepacket in the
presence of the adiabatically decreasing dc electric field (voltage) until it reaches zero at time
$t=200$~ns. By computing the squares of the final coefficients and comparing them to their initial,
$t=0$, values, the population transfer efficiency can be quantified. \autoref{fig:2} shows almost
complete population transfer (99.8~\%) from symmetric to antisymmetric rovibrational state of
ammonia.

To optimize the voltage time rise function in~\eqref{eq:1}, we have performed a series of numerical
solutions of the TDSE for different positions $z$ between $z_\text{min}$ and $z_\text{max}$ and
different values of the $a$ parameter. The results of these calculations, shown in \autoref{fig:3},
confirm that by appropriately tuning the voltage rise parameter $a$ between 0.030 and
0.070~ns$^{-1}$ one can achieve very high population transfer efficiency, more than 99.5~\%, over a
wide range of positions $z$.

\begin{figure*}
   \includegraphics[width=\linewidth]{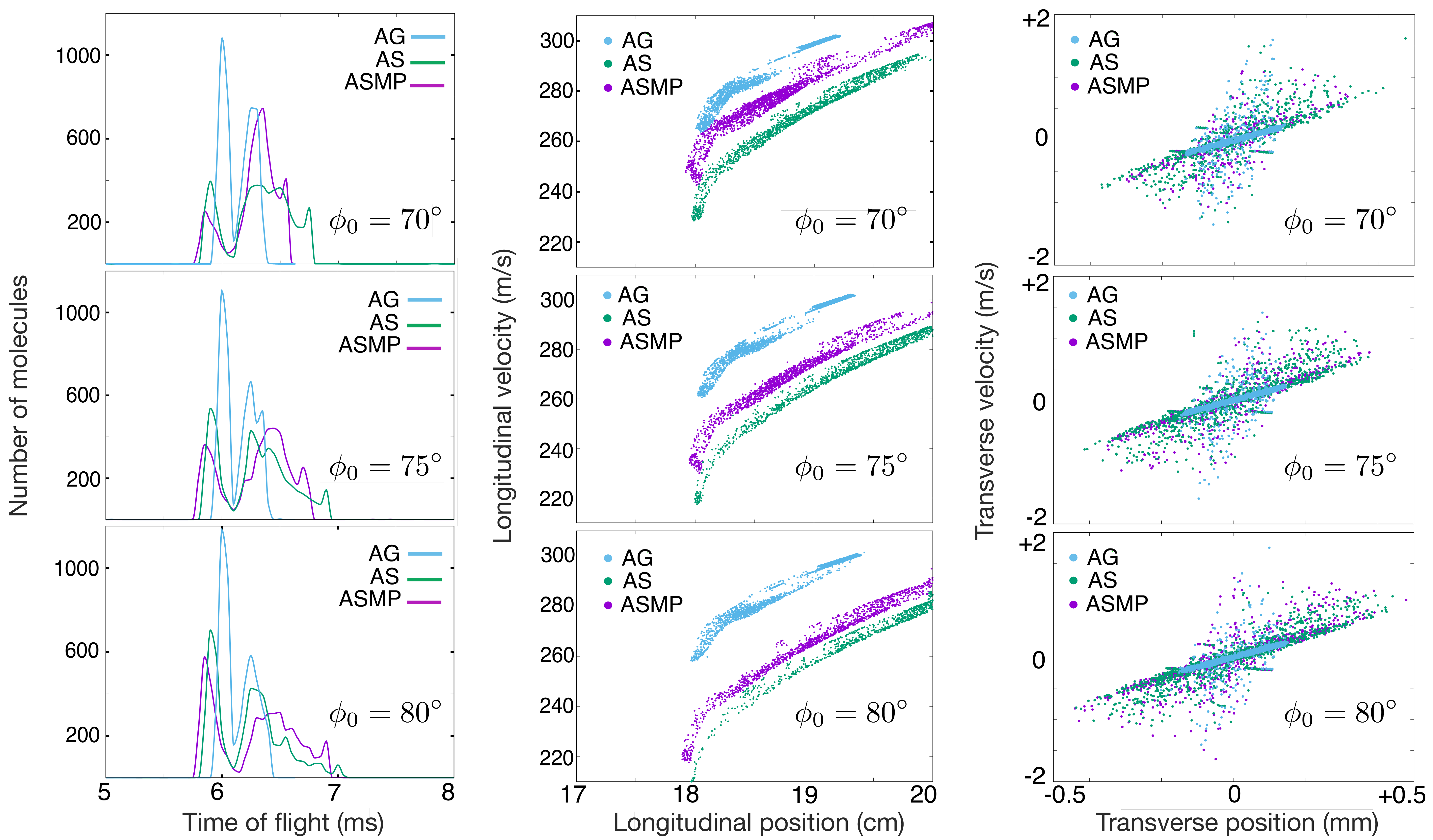}
   \caption{Calculated time-of-flight profiles and phase-space distributions of NH$_3$ for ASMP
      (magenta), AS (green) and AG (blue) deceleration schemes operating at a phase angle
      $\phi_0=70^\circ$, $75^\circ$, and $80^\circ$ with 30 deceleration stages. The transverse
      position is given by the $y$-coordinate. A $50~\mu$ bin-size was used in generating
      time-of-flight histograms.}
   \label{fig:4}
\end{figure*}

The stability of SCRAP with respect to the Stark-shift chirping rate is limited. Since the Stark
shifts of selected WFS and SFS energy levels of ammonia behave quadratically at low fields and
almost linearly at high fields, different chirping rates must be implemented for molecules located
far and near the electrodes, respectively. Optimization of the voltage-rise function has the goal of
keeping the chirping rate within the tolerance limit of the rapid-adiabatic passage for all
locations, ideally within the full acceptance volume of the decelerator.

\section{Molecular trajectory simulations}
\label{sec:traj-sim}
Classical-dynamics simulations of a prototypical Stark-deceleration experiment combined with the
alternating states method are employed to calculate molecular trajectories:
\begin{equation}
   \begin{aligned}
      m\frac{L}{\pi}\ddot{\vec{r}} = & \frac{1}{2}\left(W(t)-1\right) \left(\nabla
         E_\text{SFS}(\vec{r})-\nabla E_\text{WFS}(\vec{r})\right) \\
      & - \nabla E_\text{WFS}(\vec{r})
   \end{aligned}
   \label{eq:newton}
\end{equation}
where $\vec{r}$ is the molecule-trajectory vector, $m$ is the mass of the molecule, and
$\nabla E_\text{WFS}(\vec{r})$ and $\nabla E_\text{SFS}(\vec{r})$ are position dependent gradients
of the Stark energies of the WFS and SFS internal molecular states, respectively. $W(t)$ is a
time-dependent state occupation function with value $W(t)=1$ if the molecule is in the WFS state and
$W(t)=-1$ if it is in the SFS state. We assume that virtually complete population inversion between
WFS and SFS states can be achieved. A theoretical estimate of the losses due to incomplete
population inversion is about 0.4~\% of the number of trapped molecules per deceleration stage, see
\autoref{fig:3}.

The Stark energies and gradients for the selected WFS and SFS states of ammonia, i.e., the
inversion-split pair of $|J,k,m\rangle=|1,1,1\rangle$ states, are obtained as described above,
\autoref{sec:population}. The position-dependent gradients $\nabla E_\text{WFS}(\vec{r})$ and
$\nabla E_\text{SFS}(\vec{r})$ are calculated for the periodic electric field created by the Stark
decelerator with the electrode configuration as described in \autoref{sec:population} and the peak
voltage is set to 10~kV.

The molecule-trajectory vector $\vec{r}$ is defined by two Cartesian coordinates $x$ and $y$,
describing the transverse position of the molecule relative to the molecular beam center, and a
phase angle $\phi$ describing the relative position in the longitudinal direction, i.e., along the
molecular beam axis. Since the Stark potential is periodic over two electrode stages, the
phase-angle coordinate is defined as $\phi=\pi{}z/L$ ($\phi=0\ldots2\pi$), where $z$ is Cartesian
position along the longitudinal axis, and $L$ is the distance between centers of two neighboring
electrode stages. The values of $\phi=n\pi$ and $\phi=n\pi/2$ ($n=0,1,2$) correspond to the points
of minimum and maximum of the electric field. The voltage- and state-switching sequence is
determined by the choice of the synchronous angle $\phi_0$, a value of the phase coordinate $\phi$
for the central molecule in the beam at the moment the fields or states are switched.

We generate a uniformly-distributed molecule packet in all six dimensions in phase space, with an
average forward velocity of 300~m/s, the value typical for pulsed molecular beams~\cite{Scoles:MolBeam:1, 
Meerakker:NatPhys4:595,
	Christen:JPCA114:11189}. Initial velocity and position spreads of 5~m/s and 2~mm,
respectively, in both longitudinal and transverse directions, are reduced with respect to typical pulsed molecular beam 
experimental values, for the clarity of presentation of phase-space snapshots shown in \autoref{fig:4}. We solve 
\eqref{eq:newton} numerically for trajectories of 5000
molecules and record their final positions in phase space after 30 deceleration stages;
corresponding arrival-time and phase-space distributions are plotted in \autoref{fig:4}.

Three different voltage and internal state switching schemes are investigated. In the first scheme,
all molecules are initially prepared in the WFS state and only the dc fields in the decelerator are
switched, which is equivalent to the standard ``alternating gradient'' (AG) deceleration
method~\cite{Bethlem:JPB39:R263, Wohlfart:PRA78:033421}. In the second operation scheme all
molecules are initially in the WFS state, the decelerator's fields are kept constant using the peak
voltages, and the internal state population of molecules is switched between WFS and SFS states.
Similarly to AG, we call this method the ``alternating states'' (AS) scheme. Notably, in the AS
scheme, states are switched two times a deceleration stage at $\phi_0$ and $\phi_0+\pi/2$, i.e., at
the points close to the minimum and the maximum of electric field. The third scheme combines the
alternating states approach with the slight modulation of the electrostatic potential (ASMP), which
improves the longitudinal acceptance of the method. In this scheme, the voltage on the electrodes is
slightly decreased by 2~kV for a period of time between $\phi_0-\delta$ and $\phi_0$, that is before
the state switching, and then increased by 4~kV after the state switching, for a period of time
between $\phi_0$ and $\phi_0+\delta$, where we chose $\delta=5^\circ$.

\begin{table}[b]
   \caption{Required number of deceleration stages ($N$) for AG, AS, and ASMP deceleration schemes,
      and final molecular number densities for packets slowed down from the initial velocity
      $v = 300$~m/s to the final velocities $v = 270$~m/s and $v= 240$~m/s. Results are shown for
      two choices of synchronous angles $\phi_0$ = 70$^\circ$ and 80$^\circ$. Molecular number
      densities (\%mol) are evaluated as fractions of the initial total number of molecules (in
      percent) with the final longitudinal velocity in the range $v \pm 10$~m/s.}
   \renewcommand{\tabcolsep}{2.0mm}
   \begin{tabular}{cccccccc}
     \hline
     $v$ (m/s) & $\phi_0$ & \multicolumn{2}{c}{AG} & \multicolumn{2}{c}{AS} & \multicolumn{2}{c}{ASMP} \\
               & & $N$ & \%mol & $N$ & \%mol & $N$ & \%mol \\
     \hline
     270 & 70$^\circ$ & 27 & 26 & 14 & 51 & 18 & 46 \\
               & 80$^\circ$ & 22 & 24 & 11 & 38 & 13 & 53 \\
     240 & 70$^\circ$ & 50 & 2 & 27 & 9 & 34 & 5 \\
               & 80$^\circ$ & 42 & 1 & 21 & 5 & 24 & 10 \\
     \hline
   \end{tabular}
   \label{tab:table1}
\end{table}
\autoref{fig:4} compares the time-of-flights (TOF) and the phase space acceptance of different
schemes for three choices of the phase angle $\degree{\phi_0=70}$, \degree{75}, and \degree{80}.
The improvement of the AS and ASMP schemes is clearly visible in the TOF profiles for higher values
of the synchronous angle. For $\phi_0=80^\circ$, the ASMP peak in the TOF profile at 6.9~ms
corresponds to an average final forward velocity of 220~m/s, as shown on the longitudinal phase
plot. For comparison, using the same number of deceleration stages, the AG method slows molecules
down to an average forward velocity of 260~m/s. The ASMP works best at high values of the phase
angle, providing extra longitudinal velocity acceptance however, modulation of the potential in ASMP
results in lowered ability to slow molecules. These results are summarized in \autoref{tab:table1},
which compares the number of deceleration stages needed for the synchronous molecule to reach a
given longitudinal velocity (270 m/s and 240 m/s, respectively). The AS scheme requires
approximately half of the stages as compared to the AG scheme. The nearly doubled efficiency of the
AS technique is a direct consequence of alternating between the WFS and SFS states.

Comparison of molecular packets with equal average velocity ($v$) for the AG, AS and ASMP methods
shows that: a) the AS and ASMP schemes capture more molecules than AG, hence provide higher
molecular densities; b) at synchronous angle $\phi_0=\degree{80}$ the ASMP method gives higher
molecular densities than the AS method, i.e., 53~\% vs 38~\% at $v=270$~m/s and 10~\% vs.\ 5~\% at
$v=240$~m/s. The additional short-time voltage increase in the ASMP method provides an extra
focusing force, which when properly timed, guarantees capturing of a larger number of molecules. Further longitudinal slowing 
of molecules is possible when more deceleration stages are used. At low velocities (< 20~m/s) the phase-stability of the 
molecular packet decreases rapidly~\cite{Schnell:ACIE48:6010}, which presents a 
major challenge to 
subsequent trapping of the slowed molecules.

The transverse velocity spread shown in the right column in \autoref{fig:4} is similar for all three
methods. For $\phi_0=80^\circ$ the transverse phase-space diagrams are qualitatively identical. For
lower values of the synchronous angle, the spread in the transverse positions for AS and ASMP
methods becomes smaller than for AG method. This suggests that due to the effectively shorter
focusing and defocusing distances in AS and ASMP, large amplitudes in the transverse motion are
produced when the synchronous angle is small, which leads to larger losses of molecular density. A
compromise value of $\phi_0$ between the transverse and the longitudinal phase space acceptances for
AS and ASMP is somewhere in between \degree{70} and \degree{80}.

\section{Conclusions}
\label{sec:conclusions}
The influence of an induced inversion of the internal-state population of decelerated molecules on
the efficiency and the phase-space-acceptance characteristics of the Stark decelerator were studied.
Decelerated molecules are switched back and forth between selected WFS and SFS rovibrational states
such that the induced Stark potential is pointing upward along the molecular beam axis on both,
ascending and descending, branches of the decelerator's electric field. Specifically, we developed
and computationally tested a new scheme for population inversion in molecules, based on the rapid
adiabatic passage with a Stark-shift chirp created by controlled time-modulation of the electric
field in a decelerator. Rigorous quantum-mechanical calculations for ammonia molecule demonstrated a
population inversion efficiency larger than 99.5~\% for spatially extended, long, molecular beams.
Classical trajectory simulations demonstrated the potential of controlling the internal state of
decelerated molecules for obtaining a larger phase-space acceptance in the longitudinal direction
and more efficient deceleration.

The comprehensive internal-state control presented in this paper also gives new prospects for
separating molecules in different hyperfine states, which is important in collision studies and
high-precision spectroscopy. Calculations for these purposes can now be made with latest update to
the TROVE program~\cite{Yachmenev:JCP147:141101}, which allows for high accuracy rovibrational
calculations with nuclear-quadrupole interactions included. Finally, it is interesting to test the
present method on slow molecular beams, e.g., in combination with the recently proposed advanced
switching operation mode of the Stark decelerator~\cite{Zhang:PRA93:023408}.

\section{Acknowledgments}
We are grateful to Gabriele Santambrogio, Noah Fitch, and Mike Tarbutt for valuable discussions.
This work has been supported by the Deutsche Forschungsgemeinschaft (DFG) through the priority
program ``Quantum Dynamics in Tailored Intense Fields'' (QUTIF, SPP~1840, KU~1527/3, YA~610/1) and
through the Clusters of Excellence ``Center for Ultrafast Imaging'' (CUI, EXC~1074, ID~194651731)
and ``Advanced Imaging of Matter'' (AIM, EXC~2056, ID~390715994).

%


\begin{thebibliography}{55}%
	\makeatletter
	\providecommand \@ifxundefined [1]{%
		\@ifx{#1\undefined}
	}%
	\providecommand \@ifnum [1]{%
		\ifnum #1\expandafter \@firstoftwo
		\else \expandafter \@secondoftwo
		\fi
	}%
	\providecommand \@ifx [1]{%
		\ifx #1\expandafter \@firstoftwo
		\else \expandafter \@secondoftwo
		\fi
	}%
	\providecommand \natexlab [1]{#1}%
	\providecommand \enquote  [1]{``#1''}%
	\providecommand \bibnamefont  [1]{#1}%
	\providecommand \bibfnamefont [1]{#1}%
	\providecommand \citenamefont [1]{#1}%
	\providecommand \href@noop [0]{\@secondoftwo}%
	\providecommand \href [0]{\begingroup \@sanitize@url \@href}%
	\providecommand \@href[1]{\@@startlink{#1}\@@href}%
	\providecommand \@@href[1]{\endgroup#1\@@endlink}%
	\providecommand \@sanitize@url [0]{\catcode `\\12\catcode `\$12\catcode
		`\&12\catcode `\#12\catcode `\^12\catcode `\_12\catcode `\%12\relax}%
	\providecommand \@@startlink[1]{}%
	\providecommand \@@endlink[0]{}%
	\providecommand \url  [0]{\begingroup\@sanitize@url \@url }%
	\providecommand \@url [1]{\endgroup\@href {#1}{\urlprefix }}%
	\providecommand \urlprefix  [0]{URL }%
	\providecommand \Eprint [0]{\href }%
	\providecommand \doibase [0]{https://doi.org/}%
	\providecommand \selectlanguage [0]{\@gobble}%
	\providecommand \bibinfo  [0]{\@secondoftwo}%
	\providecommand \bibfield  [0]{\@secondoftwo}%
	\providecommand \translation [1]{[#1]}%
	\providecommand \BibitemOpen [0]{}%
	\providecommand \bibitemStop [0]{}%
	\providecommand \bibitemNoStop [0]{.\EOS\space}%
	\providecommand \EOS [0]{\spacefactor3000\relax}%
	\providecommand \BibitemShut  [1]{\csname bibitem#1\endcsname}%
	\let\auto@bib@innerbib\@empty
	\bibitem [{\citenamefont {Perreault}\ \emph {et~al.}(2020)\citenamefont
		{Perreault}, \citenamefont {Zhou}, \citenamefont {Mukherjee},\ and\
		\citenamefont {Zare}}]{Perreault:PRL124:163202}%
	\BibitemOpen
	\bibfield  {author} {\bibinfo {author} {\bibfnamefont {W.~E.}\ \bibnamefont
			{Perreault}}, \bibinfo {author} {\bibfnamefont {H.}~\bibnamefont {Zhou}},
		\bibinfo {author} {\bibfnamefont {N.}~\bibnamefont {Mukherjee}},\ and\
		\bibinfo {author} {\bibfnamefont {R.~N.}\ \bibnamefont {Zare}},\ }\bibfield
	{title} {\bibinfo {title} {Harnessing the power of adiabatic curve crossing
			to populate the highly vibrationally excited {H}$_2$ (v=7,j=0) level},\
	}\href {https://doi.org/10.1103/physrevlett.124.163202} {\bibfield  {journal}
		{\bibinfo  {journal} {Phys. Rev. Lett.}\ }\textbf {\bibinfo {volume} {124}},\
		\bibinfo {pages} {163202} (\bibinfo {year} {2020})}\BibitemShut {NoStop}%
	\bibitem [{\citenamefont {Oberst}\ \emph {et~al.}(2008)\citenamefont {Oberst},
		\citenamefont {M\"unch}, \citenamefont {Grigoryan},\ and\ \citenamefont
		{Halfmann}}]{Oberst:PRA78:033409}%
	\BibitemOpen
	\bibfield  {author} {\bibinfo {author} {\bibfnamefont {M.}~\bibnamefont
			{Oberst}}, \bibinfo {author} {\bibfnamefont {H.}~\bibnamefont {M\"unch}},
		\bibinfo {author} {\bibfnamefont {G.}~\bibnamefont {Grigoryan}},\ and\
		\bibinfo {author} {\bibfnamefont {T.}~\bibnamefont {Halfmann}},\ }\bibfield
	{title} {\bibinfo {title} {Stark-chirped rapid adiabatic passage among a
			three-state molecular system: Experimental and numerical investigations},\
	}\href {https://doi.org/10.1103/PhysRevA.78.033409} {\bibfield  {journal}
		{\bibinfo  {journal} {Phys. Rev. A}\ }\textbf {\bibinfo {volume} {78}},\
		\bibinfo {pages} {033409} (\bibinfo {year} {2008})}\BibitemShut {NoStop}%
	\bibitem [{\citenamefont {Li}\ \emph {et~al.}(2020)\citenamefont {Li},
		\citenamefont {L\"{o}tstedt}, \citenamefont {Li}, \citenamefont {Zhou},
		\citenamefont {Dong}, \citenamefont {Deng}, \citenamefont {Lu}, \citenamefont
		{Ando}, \citenamefont {Iwasaki}, \citenamefont {Fu}, \citenamefont {Wang},
		\citenamefont {Wu}, \citenamefont {Yamanouchi},\ and\ \citenamefont
		{Xu}}]{Li:PRL125:053201}%
	\BibitemOpen
	\bibfield  {author} {\bibinfo {author} {\bibfnamefont {H.}~\bibnamefont
			{Li}}, \bibinfo {author} {\bibfnamefont {E.}~\bibnamefont {L\"{o}tstedt}},
		\bibinfo {author} {\bibfnamefont {H.}~\bibnamefont {Li}}, \bibinfo {author}
		{\bibfnamefont {Y.}~\bibnamefont {Zhou}}, \bibinfo {author} {\bibfnamefont
			{N.}~\bibnamefont {Dong}}, \bibinfo {author} {\bibfnamefont {L.}~\bibnamefont
			{Deng}}, \bibinfo {author} {\bibfnamefont {P.}~\bibnamefont {Lu}}, \bibinfo
		{author} {\bibfnamefont {T.}~\bibnamefont {Ando}}, \bibinfo {author}
		{\bibfnamefont {A.}~\bibnamefont {Iwasaki}}, \bibinfo {author} {\bibfnamefont
			{Y.}~\bibnamefont {Fu}}, \bibinfo {author} {\bibfnamefont {S.}~\bibnamefont
			{Wang}}, \bibinfo {author} {\bibfnamefont {J.}~\bibnamefont {Wu}}, \bibinfo
		{author} {\bibfnamefont {K.}~\bibnamefont {Yamanouchi}},\ and\ \bibinfo
		{author} {\bibfnamefont {H.}~\bibnamefont {Xu}},\ }\bibfield  {title}
	{\bibinfo {title} {Giant enhancement of air lasing by complete population
			inversion in {N}$_2^+$},\ }\href
	{https://doi.org/10.1103/physrevlett.125.053201} {\bibfield  {journal}
		{\bibinfo  {journal} {Phys. Rev. Lett.}\ }\textbf {\bibinfo {volume} {125}},\
		\bibinfo {pages} {053201} (\bibinfo {year} {2020})}\BibitemShut {NoStop}%
	\bibitem [{\citenamefont {M{\o}ller}\ \emph {et~al.}(2008)\citenamefont
		{M{\o}ller}, \citenamefont {Madsen},\ and\ \citenamefont
		{M{\o}lmer}}]{Mller:PRL100:170504}%
	\BibitemOpen
	\bibfield  {author} {\bibinfo {author} {\bibfnamefont {D.}~\bibnamefont
			{M{\o}ller}}, \bibinfo {author} {\bibfnamefont {L.~B.}\ \bibnamefont
			{Madsen}},\ and\ \bibinfo {author} {\bibfnamefont {K.}~\bibnamefont
			{M{\o}lmer}},\ }\bibfield  {title} {\bibinfo {title} {Quantum gates and
			multiparticle entanglement by rydberg excitation blockade and adiabatic
			passage},\ }\href {https://doi.org/10.1103/physrevlett.100.170504} {\bibfield
		{journal} {\bibinfo  {journal} {Phys. Rev. Lett.}\ }\textbf {\bibinfo
			{volume} {100}},\ \bibinfo {pages} {170504} (\bibinfo {year}
		{2008})}\BibitemShut {NoStop}%
	\bibitem [{\citenamefont {Rickes}\ \emph {et~al.}(2000)\citenamefont {Rickes},
		\citenamefont {Yatsenko}, \citenamefont {Steuerwald}, \citenamefont
		{Halfmann}, \citenamefont {Shore}, \citenamefont {Vitanov},\ and\
		\citenamefont {Bergmann}}]{Rickes:JCP113:534}%
	\BibitemOpen
	\bibfield  {author} {\bibinfo {author} {\bibfnamefont {T.}~\bibnamefont
			{Rickes}}, \bibinfo {author} {\bibfnamefont {L.~P.}\ \bibnamefont
			{Yatsenko}}, \bibinfo {author} {\bibfnamefont {S.}~\bibnamefont
			{Steuerwald}}, \bibinfo {author} {\bibfnamefont {T.}~\bibnamefont
			{Halfmann}}, \bibinfo {author} {\bibfnamefont {B.~W.}\ \bibnamefont {Shore}},
		\bibinfo {author} {\bibfnamefont {N.~V.}\ \bibnamefont {Vitanov}},\ and\
		\bibinfo {author} {\bibfnamefont {K.}~\bibnamefont {Bergmann}},\ }\bibfield
	{title} {\bibinfo {title} {Efficient adiabatic population transfer by
			two-photon excitation assisted by a laser-induced stark shift},\ }\href
	{https://doi.org/10.1063/1.481829} {\bibfield  {journal} {\bibinfo  {journal}
			{J. Chem. Phys.}\ }\textbf {\bibinfo {volume} {113}},\ \bibinfo {pages} {534}
		(\bibinfo {year} {2000})}\BibitemShut {NoStop}%
	\bibitem [{\citenamefont {Rangelov}\ \emph {et~al.}(2005)\citenamefont
		{Rangelov}, \citenamefont {Vitanov}, \citenamefont {Yatsenko}, \citenamefont
		{Shore}, \citenamefont {Halfmann},\ and\ \citenamefont
		{Bergmann}}]{Rangelov:PRA72:053403}%
	\BibitemOpen
	\bibfield  {author} {\bibinfo {author} {\bibfnamefont {A.~A.}\ \bibnamefont
			{Rangelov}}, \bibinfo {author} {\bibfnamefont {N.~V.}\ \bibnamefont
			{Vitanov}}, \bibinfo {author} {\bibfnamefont {L.~P.}\ \bibnamefont
			{Yatsenko}}, \bibinfo {author} {\bibfnamefont {B.~W.}\ \bibnamefont {Shore}},
		\bibinfo {author} {\bibfnamefont {T.}~\bibnamefont {Halfmann}},\ and\
		\bibinfo {author} {\bibfnamefont {K.}~\bibnamefont {Bergmann}},\ }\bibfield
	{title} {\bibinfo {title} {Stark-shift-chirped rapid-adiabatic-passage
			technique among three states},\ }\href
	{https://doi.org/10.1103/PhysRevA.72.053403} {\bibfield  {journal} {\bibinfo
			{journal} {Phys. Rev. A}\ }\textbf {\bibinfo {volume} {72}},\ \bibinfo
		{pages} {053403} (\bibinfo {year} {2005})}\BibitemShut {NoStop}%
	\bibitem [{\citenamefont {Bethlem}\ \emph {et~al.}(1999)\citenamefont
		{Bethlem}, \citenamefont {Berden},\ and\ \citenamefont
		{Meijer}}]{Bethlem:PRL83:1558}%
	\BibitemOpen
	\bibfield  {author} {\bibinfo {author} {\bibfnamefont {H.~L.}\ \bibnamefont
			{Bethlem}}, \bibinfo {author} {\bibfnamefont {G.}~\bibnamefont {Berden}},\
		and\ \bibinfo {author} {\bibfnamefont {G.}~\bibnamefont {Meijer}},\
	}\bibfield  {title} {\bibinfo {title} {Decelerating neutral dipolar
			molecules},\ }\href {https://doi.org/10.1103/PhysRevLett.83.1558} {\bibfield
		{journal} {\bibinfo  {journal} {Phys. Rev. Lett.}\ }\textbf {\bibinfo
			{volume} {83}},\ \bibinfo {pages} {1558} (\bibinfo {year}
		{1999})}\BibitemShut {NoStop}%
	\bibitem [{\citenamefont {van~de Meerakker}\ \emph {et~al.}(2008)\citenamefont
		{van~de Meerakker}, \citenamefont {Bethlem},\ and\ \citenamefont
		{Meijer}}]{Meerakker:NatPhys4:595}%
	\BibitemOpen
	\bibfield  {author} {\bibinfo {author} {\bibfnamefont {S.~Y.~T.}\
			\bibnamefont {van~de Meerakker}}, \bibinfo {author} {\bibfnamefont {H.~L.}\
			\bibnamefont {Bethlem}},\ and\ \bibinfo {author} {\bibfnamefont
			{G.}~\bibnamefont {Meijer}},\ }\bibfield  {title} {\bibinfo {title} {Taming
			molecular beams},\ }\href {https://doi.org/10.1038/nphys1031} {\bibfield
		{journal} {\bibinfo  {journal} {Nat. Phys.}\ }\textbf {\bibinfo {volume}
			{4}},\ \bibinfo {pages} {595} (\bibinfo {year} {2008})}\BibitemShut {NoStop}%
	\bibitem [{\citenamefont {Meek}\ \emph {et~al.}(2008)\citenamefont {Meek},
		\citenamefont {Bethlem}, \citenamefont {Conrad},\ and\ \citenamefont
		{G.}}]{Meek:PRL100:153003}%
	\BibitemOpen
	\bibfield  {author} {\bibinfo {author} {\bibfnamefont {S.~A.}\ \bibnamefont
			{Meek}}, \bibinfo {author} {\bibfnamefont {H.~L.}\ \bibnamefont {Bethlem}},
		\bibinfo {author} {\bibfnamefont {H.}~\bibnamefont {Conrad}},\ and\ \bibinfo
		{author} {\bibfnamefont {M.}~\bibnamefont {G.}},\ }\bibfield  {title}
	{\bibinfo {title} {Trapping molecules on a chip in travelling potential
			wells},\ }\href@noop {} {\bibfield  {journal} {\bibinfo  {journal} {Phys.
				Rev. Lett.}\ }\textbf {\bibinfo {volume} {100}},\ \bibinfo {pages} {153003}
		(\bibinfo {year} {2008})},\ \Eprint {https://arxiv.org/abs/0801.2943}
	{arXiv:0801.2943 [physics]} \BibitemShut {NoStop}%
	\bibitem [{\citenamefont {Bell}\ and\ \citenamefont
		{Softley}(2009)}]{Bell:MP107:99}%
	\BibitemOpen
	\bibfield  {author} {\bibinfo {author} {\bibfnamefont {M.~T.}\ \bibnamefont
			{Bell}}\ and\ \bibinfo {author} {\bibfnamefont {T.~P.}\ \bibnamefont
			{Softley}},\ }\bibfield  {title} {\bibinfo {title} {Ultracold molecules and
			ultracold chemistry},\ }\href {https://doi.org/10.1080/00268970902724955}
	{\bibfield  {journal} {\bibinfo  {journal} {Mol. Phys.}\ }\textbf {\bibinfo
			{volume} {107}},\ \bibinfo {pages} {99} (\bibinfo {year} {2009})}\BibitemShut
	{NoStop}%
	\bibitem [{\citenamefont {Hogan}\ \emph {et~al.}(2009)\citenamefont {Hogan},
		\citenamefont {Seiler},\ and\ \citenamefont {Merkt}}]{Hogan:PRL103:123001}%
	\BibitemOpen
	\bibfield  {author} {\bibinfo {author} {\bibfnamefont {S.~D.}\ \bibnamefont
			{Hogan}}, \bibinfo {author} {\bibfnamefont {C.}~\bibnamefont {Seiler}},\ and\
		\bibinfo {author} {\bibfnamefont {F.}~\bibnamefont {Merkt}},\ }\bibfield
	{title} {\bibinfo {title} {Rydberg-state-enabled deceleration and trapping of
			cold molecules},\ }\href {https://doi.org/10.1103/PhysRevLett.103.123001}
	{\bibfield  {journal} {\bibinfo  {journal} {Phys. Rev. Lett.}\ }\textbf
		{\bibinfo {volume} {103}},\ \bibinfo {pages} {123001} (\bibinfo {year}
		{2009})}\BibitemShut {NoStop}%
	\bibitem [{\citenamefont {van~de Meerakker}\ \emph {et~al.}(2012)\citenamefont
		{van~de Meerakker}, \citenamefont {Bethlem}, \citenamefont {Vanhaecke},\ and\
		\citenamefont {Meijer}}]{Meerakker:CR112:4828}%
	\BibitemOpen
	\bibfield  {author} {\bibinfo {author} {\bibfnamefont {S.~Y.~T.}\
			\bibnamefont {van~de Meerakker}}, \bibinfo {author} {\bibfnamefont {H.~L.}\
			\bibnamefont {Bethlem}}, \bibinfo {author} {\bibfnamefont {N.}~\bibnamefont
			{Vanhaecke}},\ and\ \bibinfo {author} {\bibfnamefont {G.}~\bibnamefont
			{Meijer}},\ }\bibfield  {title} {\bibinfo {title} {Manipulation and control
			of molecular beams},\ }\href {https://doi.org/10.1021/cr200349r} {\bibfield
		{journal} {\bibinfo  {journal} {Chem. Rev.}\ }\textbf {\bibinfo {volume}
			{112}},\ \bibinfo {pages} {4828} (\bibinfo {year} {2012})}\BibitemShut
	{NoStop}%
	\bibitem [{\citenamefont {van Veldhoven}\ \emph {et~al.}(2004)\citenamefont
		{van Veldhoven}, \citenamefont {K{\"u}pper}, \citenamefont {Bethlem},
		\citenamefont {Sartakov}, \citenamefont {van Roij},\ and\ \citenamefont
		{Meijer}}]{Veldhoven:EPJD31:337}%
	\BibitemOpen
	\bibfield  {author} {\bibinfo {author} {\bibfnamefont {J.}~\bibnamefont {van
				Veldhoven}}, \bibinfo {author} {\bibfnamefont {J.}~\bibnamefont
			{K{\"u}pper}}, \bibinfo {author} {\bibfnamefont {H.~L.}\ \bibnamefont
			{Bethlem}}, \bibinfo {author} {\bibfnamefont {B.}~\bibnamefont {Sartakov}},
		\bibinfo {author} {\bibfnamefont {A.~J.~A.}\ \bibnamefont {van Roij}},\ and\
		\bibinfo {author} {\bibfnamefont {G.}~\bibnamefont {Meijer}},\ }\bibfield
	{title} {\bibinfo {title} {Decelerated molecular beams for high-resolution
			spectroscopy: The hyperfine structure of {$^{15}${ND}$_3$}},\ }\href
	{https://doi.org/10.1140/epjd/e2004-00160-9} {\bibfield  {journal} {\bibinfo
			{journal} {Eur. Phys. J. D}\ }\textbf {\bibinfo {volume} {31}},\ \bibinfo
		{pages} {337} (\bibinfo {year} {2004})}\BibitemShut {NoStop}%
	\bibitem [{\citenamefont {Krems}(2005)}]{Krems:IRPC24:99}%
	\BibitemOpen
	\bibfield  {author} {\bibinfo {author} {\bibfnamefont {R.~V.}\ \bibnamefont
			{Krems}},\ }\bibfield  {title} {\bibinfo {title} {Molecules near absolute
			zero and external field control of atomic and molecular dynamics},\ }\href
	{https://doi.org/10.1080/01442350500167161} {\bibfield  {journal} {\bibinfo
			{journal} {Int. Rev. Phys. Chem.}\ }\textbf {\bibinfo {volume} {24}},\
		\bibinfo {pages} {99} (\bibinfo {year} {2005})}\BibitemShut {NoStop}%
	\bibitem [{\citenamefont {Gilijamse}\ \emph {et~al.}(2006)\citenamefont
		{Gilijamse}, \citenamefont {Hoekstra}, \citenamefont {van~de Meerakker},
		\citenamefont {Groenenboom},\ and\ \citenamefont
		{Meijer}}]{Gilijamse:Science313:1617}%
	\BibitemOpen
	\bibfield  {author} {\bibinfo {author} {\bibfnamefont {J.~J.}\ \bibnamefont
			{Gilijamse}}, \bibinfo {author} {\bibfnamefont {S.}~\bibnamefont {Hoekstra}},
		\bibinfo {author} {\bibfnamefont {S.~Y.~T.}\ \bibnamefont {van~de
				Meerakker}}, \bibinfo {author} {\bibfnamefont {G.~C.}\ \bibnamefont
			{Groenenboom}},\ and\ \bibinfo {author} {\bibfnamefont {G.}~\bibnamefont
			{Meijer}},\ }\bibfield  {title} {\bibinfo {title} {Near-threshold inelastic
			collisions using molecular beams with a tunable velocity},\ }\href
	{https://doi.org/10.1126/science.1131867} {\bibfield  {journal} {\bibinfo
			{journal} {Science}\ }\textbf {\bibinfo {volume} {313}},\ \bibinfo {pages}
		{1617} (\bibinfo {year} {2006})}\BibitemShut {NoStop}%
	\bibitem [{\citenamefont {Hudson}\ \emph {et~al.}(2006)\citenamefont {Hudson},
		\citenamefont {Lewandowski}, \citenamefont {Sawyer},\ and\ \citenamefont
		{Ye}}]{Hudson:PRL96:143004}%
	\BibitemOpen
	\bibfield  {author} {\bibinfo {author} {\bibfnamefont {E.~R.}\ \bibnamefont
			{Hudson}}, \bibinfo {author} {\bibfnamefont {H.~J.}\ \bibnamefont
			{Lewandowski}}, \bibinfo {author} {\bibfnamefont {B.~C.}\ \bibnamefont
			{Sawyer}},\ and\ \bibinfo {author} {\bibfnamefont {J.}~\bibnamefont {Ye}},\
	}\bibfield  {title} {\bibinfo {title} {Cold molecule spectroscopy for
			constraining the evolution of the fine structure constant},\ }\href
	{http://link.aps.org/abstract/PRL/v96/e143004} {\bibfield  {journal}
		{\bibinfo  {journal} {Phys. Rev. Lett.}\ }\textbf {\bibinfo {volume} {96}},\
		\bibinfo {pages} {143004} (\bibinfo {year} {2006})}\BibitemShut {NoStop}%
	\bibitem [{\citenamefont {Schnell}\ and\ \citenamefont
		{K{\"u}pper}(2011)}]{Schnell:FD150:33}%
	\BibitemOpen
	\bibfield  {author} {\bibinfo {author} {\bibfnamefont {M.}~\bibnamefont
			{Schnell}}\ and\ \bibinfo {author} {\bibfnamefont {J.}~\bibnamefont
			{K{\"u}pper}},\ }\bibfield  {title} {\bibinfo {title} {Tailored molecular
			samples for precision spectroscopy experiments},\ }\href
	{https://doi.org/10.1039/c0fd00009d} {\bibfield  {journal} {\bibinfo
			{journal} {Faraday Disc.}\ }\textbf {\bibinfo {volume} {150}},\ \bibinfo
		{pages} {33} (\bibinfo {year} {2011})}\BibitemShut {NoStop}%
	\bibitem [{\citenamefont {Stuhl}\ \emph {et~al.}(2014)\citenamefont {Stuhl},
		\citenamefont {Hummon},\ and\ \citenamefont {Ye}}]{Stuhl:ARPC65:501}%
	\BibitemOpen
	\bibfield  {author} {\bibinfo {author} {\bibfnamefont {B.~K.}\ \bibnamefont
			{Stuhl}}, \bibinfo {author} {\bibfnamefont {M.~T.}\ \bibnamefont {Hummon}},\
		and\ \bibinfo {author} {\bibfnamefont {J.}~\bibnamefont {Ye}},\ }\bibfield
	{title} {\bibinfo {title} {Cold state-selected molecular collisions and
			reactions},\ }\href {https://doi.org/10.1146/annurev-physchem-040513-103744}
	{\bibfield  {journal} {\bibinfo  {journal} {Annu. Rev. Phys. Chem.}\ }\textbf
		{\bibinfo {volume} {65}},\ \bibinfo {pages} {501} (\bibinfo {year}
		{2014})}\BibitemShut {NoStop}%
	\bibitem [{\citenamefont {Wall}\ \emph {et~al.}(2015)\citenamefont {Wall},
		\citenamefont {Maeda},\ and\ \citenamefont {Carr}}]{Wall:NJP17:025001}%
	\BibitemOpen
	\bibfield  {author} {\bibinfo {author} {\bibfnamefont {M.~L.}\ \bibnamefont
			{Wall}}, \bibinfo {author} {\bibfnamefont {K.}~\bibnamefont {Maeda}},\ and\
		\bibinfo {author} {\bibfnamefont {L.~D.}\ \bibnamefont {Carr}},\ }\bibfield
	{title} {\bibinfo {title} {Realizing unconventional quantum magnetism with
			symmetric top molecules},\ }\href
	{https://doi.org/10.1088/1367-2630/17/2/025001} {\bibfield  {journal}
		{\bibinfo  {journal} {New J. Phys.}\ }\textbf {\bibinfo {volume} {17}},\
		\bibinfo {pages} {025001} (\bibinfo {year} {2015})}\BibitemShut {NoStop}%
	\bibitem [{\citenamefont {Vogels}\ \emph {et~al.}(2015)\citenamefont {Vogels},
		\citenamefont {Onvlee}, \citenamefont {Chefdeville}, \citenamefont {van~der
			Avoird}, \citenamefont {Groenenboom},\ and\ \citenamefont
		{Meerakker}}]{Vogels:Science350:787}%
	\BibitemOpen
	\bibfield  {author} {\bibinfo {author} {\bibfnamefont {S.~N.}\ \bibnamefont
			{Vogels}}, \bibinfo {author} {\bibfnamefont {J.}~\bibnamefont {Onvlee}},
		\bibinfo {author} {\bibfnamefont {S.}~\bibnamefont {Chefdeville}}, \bibinfo
		{author} {\bibfnamefont {A.}~\bibnamefont {van~der Avoird}}, \bibinfo
		{author} {\bibfnamefont {G.~C.}\ \bibnamefont {Groenenboom}},\ and\ \bibinfo
		{author} {\bibfnamefont {S.~Y.~T.}\ \bibnamefont {Meerakker}},\ }\bibfield
	{title} {\bibinfo {title} {Imaging resonances in low-energy no-he inelastic
			collisions.},\ }\href {https://doi.org/10.1126/science.aad2356} {\bibfield
		{journal} {\bibinfo  {journal} {Science}\ }\textbf {\bibinfo {volume}
			{350}},\ \bibinfo {pages} {787} (\bibinfo {year} {2015})},\ \Eprint
	{https://arxiv.org/abs/1510.00299} {arXiv:1510.00299 [physics]} \BibitemShut
	{NoStop}%
	\bibitem [{\citenamefont {Auerbach}\ \emph {et~al.}(1966)\citenamefont
		{Auerbach}, \citenamefont {Bromberg},\ and\ \citenamefont
		{Wharton}}]{Auerbach:JCP45:2160}%
	\BibitemOpen
	\bibfield  {author} {\bibinfo {author} {\bibfnamefont {D.}~\bibnamefont
			{Auerbach}}, \bibinfo {author} {\bibfnamefont {E.~E.~A.}\ \bibnamefont
			{Bromberg}},\ and\ \bibinfo {author} {\bibfnamefont {L.}~\bibnamefont
			{Wharton}},\ }\bibfield  {title} {\bibinfo {title} {Alternate-gradient
			focusing of molecular beams},\ }\href {https://doi.org/10.1063/1.1727902}
	{\bibfield  {journal} {\bibinfo  {journal} {J. Chem. Phys.}\ }\textbf
		{\bibinfo {volume} {45}},\ \bibinfo {pages} {2160} (\bibinfo {year}
		{1966})}\BibitemShut {NoStop}%
	\bibitem [{\citenamefont {Bethlem}\ \emph
		{et~al.}(2002{\natexlab{a}})\citenamefont {Bethlem}, \citenamefont {van
			Roij}, \citenamefont {Jongma},\ and\ \citenamefont
		{Meijer}}]{Bethlem:PRL88:133003}%
	\BibitemOpen
	\bibfield  {author} {\bibinfo {author} {\bibfnamefont {H.~L.}\ \bibnamefont
			{Bethlem}}, \bibinfo {author} {\bibfnamefont {A.~J.~A.}\ \bibnamefont {van
				Roij}}, \bibinfo {author} {\bibfnamefont {R.~T.}\ \bibnamefont {Jongma}},\
		and\ \bibinfo {author} {\bibfnamefont {G.}~\bibnamefont {Meijer}},\
	}\bibfield  {title} {\bibinfo {title} {Alternate gradient focusing and
			deceleration of a molecular beam},\ }\href
	{https://doi.org/10.1103/PhysRevLett.88.133003} {\bibfield  {journal}
		{\bibinfo  {journal} {Phys. Rev. Lett.}\ }\textbf {\bibinfo {volume} {88}},\
		\bibinfo {pages} {133003} (\bibinfo {year} {2002}{\natexlab{a}})}\BibitemShut
	{NoStop}%
	\bibitem [{\citenamefont {Tarbutt}\ \emph {et~al.}(2004)\citenamefont
		{Tarbutt}, \citenamefont {Bethlem}, \citenamefont {Hudson}, \citenamefont
		{Ryabov}, \citenamefont {Ryzhov}, \citenamefont {Sauer}, \citenamefont
		{Meijer},\ and\ \citenamefont {Hinds}}]{Tarbutt:PRL92:173002}%
	\BibitemOpen
	\bibfield  {author} {\bibinfo {author} {\bibfnamefont {M.~R.}\ \bibnamefont
			{Tarbutt}}, \bibinfo {author} {\bibfnamefont {H.~L.}\ \bibnamefont
			{Bethlem}}, \bibinfo {author} {\bibfnamefont {J.~J.}\ \bibnamefont {Hudson}},
		\bibinfo {author} {\bibfnamefont {V.~L.}\ \bibnamefont {Ryabov}}, \bibinfo
		{author} {\bibfnamefont {V.~A.}\ \bibnamefont {Ryzhov}}, \bibinfo {author}
		{\bibfnamefont {B.~E.}\ \bibnamefont {Sauer}}, \bibinfo {author}
		{\bibfnamefont {G.}~\bibnamefont {Meijer}},\ and\ \bibinfo {author}
		{\bibfnamefont {E.~A.}\ \bibnamefont {Hinds}},\ }\bibfield  {title} {\bibinfo
		{title} {Slowing heavy, ground-state molecules using an alternating gradient
			decelerator},\ }\href {https://doi.org/10.1103/PhysRevLett.92.173002}
	{\bibfield  {journal} {\bibinfo  {journal} {Phys. Rev. Lett.}\ }\textbf
		{\bibinfo {volume} {92}},\ \bibinfo {pages} {173002} (\bibinfo {year}
		{2004})}\BibitemShut {NoStop}%
	\bibitem [{\citenamefont {Bethlem}\ \emph {et~al.}(2006)\citenamefont
		{Bethlem}, \citenamefont {Tarbutt}, \citenamefont {K{\"u}pper}, \citenamefont
		{Carty}, \citenamefont {Wohlfart}, \citenamefont {Hinds},\ and\ \citenamefont
		{Meijer}}]{Bethlem:JPB39:R263}%
	\BibitemOpen
	\bibfield  {author} {\bibinfo {author} {\bibfnamefont {H.~L.}\ \bibnamefont
			{Bethlem}}, \bibinfo {author} {\bibfnamefont {M.~R.}\ \bibnamefont
			{Tarbutt}}, \bibinfo {author} {\bibfnamefont {J.}~\bibnamefont {K{\"u}pper}},
		\bibinfo {author} {\bibfnamefont {D.}~\bibnamefont {Carty}}, \bibinfo
		{author} {\bibfnamefont {K.}~\bibnamefont {Wohlfart}}, \bibinfo {author}
		{\bibfnamefont {E.~A.}\ \bibnamefont {Hinds}},\ and\ \bibinfo {author}
		{\bibfnamefont {G.}~\bibnamefont {Meijer}},\ }\bibfield  {title} {\bibinfo
		{title} {Alternating gradient focusing and deceleration of polar molecules},\
	}\href {https://doi.org/10.1088/0953-4075/39/16/R01} {\bibfield  {journal}
		{\bibinfo  {journal} {J. Phys. B}\ }\textbf {\bibinfo {volume} {39}},\
		\bibinfo {pages} {R263} (\bibinfo {year} {2006})},\ \Eprint
	{https://arxiv.org/abs/0604020} {arXiv:0604020 [physics]} \BibitemShut
	{NoStop}%
	\bibitem [{\citenamefont {Wohlfart}\ \emph
		{et~al.}(2008{\natexlab{a}})\citenamefont {Wohlfart}, \citenamefont
		{Gr{\"a}tz}, \citenamefont {Filsinger}, \citenamefont {Haak}, \citenamefont
		{Meijer},\ and\ \citenamefont {K{\"u}pper}}]{Wohlfart:PRA77:031404R}%
	\BibitemOpen
	\bibfield  {author} {\bibinfo {author} {\bibfnamefont {K.}~\bibnamefont
			{Wohlfart}}, \bibinfo {author} {\bibfnamefont {F.}~\bibnamefont {Gr{\"a}tz}},
		\bibinfo {author} {\bibfnamefont {F.}~\bibnamefont {Filsinger}}, \bibinfo
		{author} {\bibfnamefont {H.}~\bibnamefont {Haak}}, \bibinfo {author}
		{\bibfnamefont {G.}~\bibnamefont {Meijer}},\ and\ \bibinfo {author}
		{\bibfnamefont {J.}~\bibnamefont {K{\"u}pper}},\ }\bibfield  {title}
	{\bibinfo {title} {Alternating-gradient focusing and deceleration of large
			molecules},\ }\href {https://doi.org/10.1103/PhysRevA.77.031404} {\bibfield
		{journal} {\bibinfo  {journal} {Phys. Rev. A}\ }\textbf {\bibinfo {volume}
			{77}},\ \bibinfo {pages} {031404(R)} (\bibinfo {year}
		{2008}{\natexlab{a}})},\ \Eprint {https://arxiv.org/abs/0803.0650}
	{arXiv:0803.0650 [physics]} \BibitemShut {NoStop}%
	\bibitem [{\citenamefont {Wohlfart}\ \emph
		{et~al.}(2008{\natexlab{b}})\citenamefont {Wohlfart}, \citenamefont
		{Filsinger}, \citenamefont {Gr{\"a}tz}, \citenamefont {K{\"u}pper},\ and\
		\citenamefont {Meijer}}]{Wohlfart:PRA78:033421}%
	\BibitemOpen
	\bibfield  {author} {\bibinfo {author} {\bibfnamefont {K.}~\bibnamefont
			{Wohlfart}}, \bibinfo {author} {\bibfnamefont {F.}~\bibnamefont {Filsinger}},
		\bibinfo {author} {\bibfnamefont {F.}~\bibnamefont {Gr{\"a}tz}}, \bibinfo
		{author} {\bibfnamefont {J.}~\bibnamefont {K{\"u}pper}},\ and\ \bibinfo
		{author} {\bibfnamefont {G.}~\bibnamefont {Meijer}},\ }\bibfield  {title}
	{\bibinfo {title} {Stark deceleration of {OH} radicals in low-field-seeking
			and high-field-seeking quantum states},\ }\href
	{https://doi.org/10.1103/PhysRevA.78.033421} {\bibfield  {journal} {\bibinfo
			{journal} {Phys. Rev. A}\ }\textbf {\bibinfo {volume} {78}},\ \bibinfo
		{pages} {033421} (\bibinfo {year} {2008}{\natexlab{b}})}\BibitemShut
	{NoStop}%
	\bibitem [{\citenamefont {Filsinger}\ \emph {et~al.}(2008)\citenamefont
		{Filsinger}, \citenamefont {Erlekam}, \citenamefont {von Helden},
		\citenamefont {K{\"u}pper},\ and\ \citenamefont
		{Meijer}}]{Filsinger:PRL100:133003}%
	\BibitemOpen
	\bibfield  {author} {\bibinfo {author} {\bibfnamefont {F.}~\bibnamefont
			{Filsinger}}, \bibinfo {author} {\bibfnamefont {U.}~\bibnamefont {Erlekam}},
		\bibinfo {author} {\bibfnamefont {G.}~\bibnamefont {von Helden}}, \bibinfo
		{author} {\bibfnamefont {J.}~\bibnamefont {K{\"u}pper}},\ and\ \bibinfo
		{author} {\bibfnamefont {G.}~\bibnamefont {Meijer}},\ }\bibfield  {title}
	{\bibinfo {title} {Selector for structural isomers of neutral molecules},\
	}\href {https://doi.org/10.1103/PhysRevLett.100.133003} {\bibfield  {journal}
		{\bibinfo  {journal} {Phys. Rev. Lett.}\ }\textbf {\bibinfo {volume} {100}},\
		\bibinfo {pages} {133003} (\bibinfo {year} {2008})},\ \Eprint
	{https://arxiv.org/abs/0802.2795} {arXiv:0802.2795 [physics]} \BibitemShut
	{NoStop}%
	\bibitem [{\citenamefont {Chang}\ \emph {et~al.}(2015)\citenamefont {Chang},
		\citenamefont {Horke}, \citenamefont {Trippel},\ and\ \citenamefont
		{Küpper}}]{Chang:IRPC34:557}%
	\BibitemOpen
	\bibfield  {author} {\bibinfo {author} {\bibfnamefont {Y.-P.}\ \bibnamefont
			{Chang}}, \bibinfo {author} {\bibfnamefont {D.~A.}\ \bibnamefont {Horke}},
		\bibinfo {author} {\bibfnamefont {S.}~\bibnamefont {Trippel}},\ and\ \bibinfo
		{author} {\bibfnamefont {J.}~\bibnamefont {Küpper}},\ }\bibfield  {title}
	{\bibinfo {title} {Spatially-controlled complex molecules and their
			applications},\ }\href {https://doi.org/10.1080/0144235X.2015.1077838}
	{\bibfield  {journal} {\bibinfo  {journal} {Int. Rev. Phys. Chem.}\ }\textbf
		{\bibinfo {volume} {34}},\ \bibinfo {pages} {557} (\bibinfo {year} {2015})},\
	\Eprint {https://arxiv.org/abs/1505.05632} {arXiv:1505.05632 [physics]}
	\BibitemShut {NoStop}%
	\bibitem [{\citenamefont {Hudson}(2009)}]{Hudson:PRA79:061407}%
	\BibitemOpen
	\bibfield  {author} {\bibinfo {author} {\bibfnamefont {E.~R.}\ \bibnamefont
			{Hudson}},\ }\bibfield  {title} {\bibinfo {title} {Deceleration of continuous
			molecular beams},\ }\href {https://doi.org/10.1103/PhysRevA.79.061407}
	{\bibfield  {journal} {\bibinfo  {journal} {Phys. Rev. A}\ }\textbf {\bibinfo
			{volume} {79}},\ \bibinfo {pages} {061407} (\bibinfo {year}
		{2009})}\BibitemShut {NoStop}%
	\bibitem [{\citenamefont {Zeppenfeld}\ \emph {et~al.}(2012)\citenamefont
		{Zeppenfeld}, \citenamefont {Englert}, \citenamefont {Gl\"{o}ckner},
		\citenamefont {Prehn}, \citenamefont {Mielenz}, \citenamefont {Sommer},
		\citenamefont {van Buuren}, \citenamefont {Motsch},\ and\ \citenamefont
		{Rempe}}]{Zeppenfeld:Nature491:570}%
	\BibitemOpen
	\bibfield  {author} {\bibinfo {author} {\bibfnamefont {M.}~\bibnamefont
			{Zeppenfeld}}, \bibinfo {author} {\bibfnamefont {B.~G.~U.}\ \bibnamefont
			{Englert}}, \bibinfo {author} {\bibfnamefont {R.}~\bibnamefont
			{Gl\"{o}ckner}}, \bibinfo {author} {\bibfnamefont {A.}~\bibnamefont {Prehn}},
		\bibinfo {author} {\bibfnamefont {M.}~\bibnamefont {Mielenz}}, \bibinfo
		{author} {\bibfnamefont {C.}~\bibnamefont {Sommer}}, \bibinfo {author}
		{\bibfnamefont {L.~D.}\ \bibnamefont {van Buuren}}, \bibinfo {author}
		{\bibfnamefont {M.}~\bibnamefont {Motsch}},\ and\ \bibinfo {author}
		{\bibfnamefont {G.}~\bibnamefont {Rempe}},\ }\bibfield  {title} {\bibinfo
		{title} {Sisyphus cooling of electrically trapped polyatomic molecules},\
	}\href {https://doi.org/10.1038/nature11595} {\bibfield  {journal} {\bibinfo
			{journal} {Nature}\ }\textbf {\bibinfo {volume} {491}},\ \bibinfo {pages}
		{570} (\bibinfo {year} {2012})},\ \Eprint {https://arxiv.org/abs/1208.0046}
	{arXiv:1208.0046 [physics]} \BibitemShut {NoStop}%
	\bibitem [{\citenamefont {Prehn}\ \emph {et~al.}(2016)\citenamefont {Prehn},
		\citenamefont {Ibr\"ugger}, \citenamefont {Gl\"ockner}, \citenamefont
		{Rempe},\ and\ \citenamefont {Zeppenfeld}}]{Prehn:PRL116:063005}%
	\BibitemOpen
	\bibfield  {author} {\bibinfo {author} {\bibfnamefont {A.}~\bibnamefont
			{Prehn}}, \bibinfo {author} {\bibfnamefont {M.}~\bibnamefont {Ibr\"ugger}},
		\bibinfo {author} {\bibfnamefont {R.}~\bibnamefont {Gl\"ockner}}, \bibinfo
		{author} {\bibfnamefont {G.}~\bibnamefont {Rempe}},\ and\ \bibinfo {author}
		{\bibfnamefont {M.}~\bibnamefont {Zeppenfeld}},\ }\bibfield  {title}
	{\bibinfo {title} {Optoelectrical cooling of polar molecules to
			submillikelvin temperatures},\ }\href
	{https://doi.org/10.1103/PhysRevLett.116.063005} {\bibfield  {journal}
		{\bibinfo  {journal} {Phys. Rev. Lett.}\ }\textbf {\bibinfo {volume} {116}},\
		\bibinfo {pages} {063005} (\bibinfo {year} {2016})},\ \Eprint
	{https://arxiv.org/abs/1511.09427} {arXiv:1511.09427 [physics]} \BibitemShut
	{NoStop}%
	\bibitem [{\citenamefont {Mukherjee}\ \emph {et~al.}(2020)\citenamefont
		{Mukherjee}, \citenamefont {Widhalm}, \citenamefont {Siebert}, \citenamefont
		{Krehs}, \citenamefont {Sharma}, \citenamefont {Thiede}, \citenamefont
		{Reuter}, \citenamefont {F\"{o}rstner},\ and\ \citenamefont
		{Zrenner}}]{Mukherjee:APL116:251103}%
	\BibitemOpen
	\bibfield  {author} {\bibinfo {author} {\bibfnamefont {A.}~\bibnamefont
			{Mukherjee}}, \bibinfo {author} {\bibfnamefont {A.}~\bibnamefont {Widhalm}},
		\bibinfo {author} {\bibfnamefont {D.}~\bibnamefont {Siebert}}, \bibinfo
		{author} {\bibfnamefont {S.}~\bibnamefont {Krehs}}, \bibinfo {author}
		{\bibfnamefont {N.}~\bibnamefont {Sharma}}, \bibinfo {author} {\bibfnamefont
			{A.}~\bibnamefont {Thiede}}, \bibinfo {author} {\bibfnamefont
			{D.}~\bibnamefont {Reuter}}, \bibinfo {author} {\bibfnamefont
			{J.}~\bibnamefont {F\"{o}rstner}},\ and\ \bibinfo {author} {\bibfnamefont
			{A.}~\bibnamefont {Zrenner}},\ }\bibfield  {title} {\bibinfo {title}
		{Electrically controlled rapid adiabatic passage in a single quantum dot},\
	}\href {https://doi.org/10.1063/5.0012257} {\bibfield  {journal} {\bibinfo
			{journal} {Applied Physics Letters}\ }\textbf {\bibinfo {volume} {116}},\
		\bibinfo {pages} {251103} (\bibinfo {year} {2020})}\BibitemShut {NoStop}%
	\bibitem [{\citenamefont {Hutzler}\ \emph {et~al.}(2012)\citenamefont
		{Hutzler}, \citenamefont {Lu},\ and\ \citenamefont
		{Doyle}}]{Hutzler:CR112:4803}%
	\BibitemOpen
	\bibfield  {author} {\bibinfo {author} {\bibfnamefont {N.~R.}\ \bibnamefont
			{Hutzler}}, \bibinfo {author} {\bibfnamefont {H.-I.}\ \bibnamefont {Lu}},\
		and\ \bibinfo {author} {\bibfnamefont {J.~M.}\ \bibnamefont {Doyle}},\
	}\bibfield  {title} {\bibinfo {title} {The buffer gas beam: An intense, cold,
			and slow source for atoms and molecules},\ }\href
	{https://doi.org/10.1021/cr200362u} {\bibfield  {journal} {\bibinfo
			{journal} {Chem. Rev.}\ }\textbf {\bibinfo {volume} {112}},\ \bibinfo {pages}
		{4803} (\bibinfo {year} {2012})},\ \Eprint {https://arxiv.org/abs/1111.2841}
	{arXiv:1111.2841 [physics]} \BibitemShut {NoStop}%
	\bibitem [{\citenamefont {Singh}\ \emph {et~al.}(2018)\citenamefont {Singh},
		\citenamefont {Samanta}, \citenamefont {Roth}, \citenamefont {Gusa},
		\citenamefont {Ossenbr{\"u}ggen}, \citenamefont {Rubinsky}, \citenamefont
		{Horke},\ and\ \citenamefont {K{\"u}pper}}]{Singh:PRA97:032704}%
	\BibitemOpen
	\bibfield  {author} {\bibinfo {author} {\bibfnamefont {V.}~\bibnamefont
			{Singh}}, \bibinfo {author} {\bibfnamefont {A.~K.}\ \bibnamefont {Samanta}},
		\bibinfo {author} {\bibfnamefont {N.}~\bibnamefont {Roth}}, \bibinfo {author}
		{\bibfnamefont {D.}~\bibnamefont {Gusa}}, \bibinfo {author} {\bibfnamefont
			{T.}~\bibnamefont {Ossenbr{\"u}ggen}}, \bibinfo {author} {\bibfnamefont
			{I.}~\bibnamefont {Rubinsky}}, \bibinfo {author} {\bibfnamefont {D.~A.}\
			\bibnamefont {Horke}},\ and\ \bibinfo {author} {\bibfnamefont
			{J.}~\bibnamefont {K{\"u}pper}},\ }\bibfield  {title} {\bibinfo {title}
		{Optimized cell geometry for buffer-gas-cooled molecular-beam sources},\
	}\href {https://doi.org/10.1103/PhysRevA.97.032704} {\bibfield  {journal}
		{\bibinfo  {journal} {Phys. Rev. A}\ }\textbf {\bibinfo {volume} {97}},\
		\bibinfo {pages} {032704} (\bibinfo {year} {2018})},\ \Eprint
	{https://arxiv.org/abs/1801.10586} {arXiv:1801.10586 [physics]} \BibitemShut
	{NoStop}%
	\bibitem [{\citenamefont {Barry}\ \emph {et~al.}(2012)\citenamefont {Barry},
		\citenamefont {Shuman}, \citenamefont {Norrgard},\ and\ \citenamefont
		{DeMille}}]{Barry:PRL108:103002}%
	\BibitemOpen
	\bibfield  {author} {\bibinfo {author} {\bibfnamefont {J.}~\bibnamefont
			{Barry}}, \bibinfo {author} {\bibfnamefont {E.}~\bibnamefont {Shuman}},
		\bibinfo {author} {\bibfnamefont {E.}~\bibnamefont {Norrgard}},\ and\
		\bibinfo {author} {\bibfnamefont {D.}~\bibnamefont {DeMille}},\ }\bibfield
	{title} {\bibinfo {title} {Laser radiation pressure slowing of a molecular
			beam},\ }\href {https://doi.org/10.1103/PhysRevLett.108.103002} {\bibfield
		{journal} {\bibinfo  {journal} {Phys. Rev. Lett.}\ }\textbf {\bibinfo
			{volume} {108}},\ \bibinfo {pages} {103002} (\bibinfo {year} {2012})},\
	\Eprint {https://arxiv.org/abs/1110.4890} {arXiv:1110.4890 [physics]}
	\BibitemShut {NoStop}%
	\bibitem [{\citenamefont {Truppe}\ \emph {et~al.}(2017)\citenamefont {Truppe},
		\citenamefont {Williams}, \citenamefont {Fitch}, \citenamefont {Hambach},
		\citenamefont {Wall}, \citenamefont {Hinds}, \citenamefont {Sauer},\ and\
		\citenamefont {Tarbutt}}]{Truppe:NJP19:022001}%
	\BibitemOpen
	\bibfield  {author} {\bibinfo {author} {\bibfnamefont {S.}~\bibnamefont
			{Truppe}}, \bibinfo {author} {\bibfnamefont {H.~J.}\ \bibnamefont
			{Williams}}, \bibinfo {author} {\bibfnamefont {N.~J.}\ \bibnamefont {Fitch}},
		\bibinfo {author} {\bibfnamefont {M.}~\bibnamefont {Hambach}}, \bibinfo
		{author} {\bibfnamefont {T.~E.}\ \bibnamefont {Wall}}, \bibinfo {author}
		{\bibfnamefont {E.~A.}\ \bibnamefont {Hinds}}, \bibinfo {author}
		{\bibfnamefont {B.~E.}\ \bibnamefont {Sauer}},\ and\ \bibinfo {author}
		{\bibfnamefont {M.~R.}\ \bibnamefont {Tarbutt}},\ }\bibfield  {title}
	{\bibinfo {title} {An intense, cold, velocity-controlled molecular beam by
			frequency-chirped laser slowing},\ }\href
	{https://doi.org/10.1088/1367-2630/aa5ca2} {\bibfield  {journal} {\bibinfo
			{journal} {New J. Phys.}\ }\textbf {\bibinfo {volume} {19}},\ \bibinfo
		{pages} {022001} (\bibinfo {year} {2017})},\ \Eprint
	{https://arxiv.org/abs/1605.06055} {arXiv:1605.06055 [physics]} \BibitemShut
	{NoStop}%
	\bibitem [{\citenamefont {Comparat}(2014)}]{Comparat:PRA89:043410}%
	\BibitemOpen
	\bibfield  {author} {\bibinfo {author} {\bibfnamefont {D.}~\bibnamefont
			{Comparat}},\ }\bibfield  {title} {\bibinfo {title} {Molecular cooling via
			sisyphus processes},\ }\href {https://doi.org/10.1103/PhysRevA.89.043410}
	{\bibfield  {journal} {\bibinfo  {journal} {Phys. Rev. A}\ }\textbf {\bibinfo
			{volume} {89}},\ \bibinfo {pages} {043410} (\bibinfo {year} {2014})},\
	\Eprint {https://arxiv.org/abs/1404.2689} {arXiv:1404.2689 [physics]}
	\BibitemShut {NoStop}%
	\bibitem [{\citenamefont {Lim}\ \emph {et~al.}(2015)\citenamefont {Lim},
		\citenamefont {Frye}, \citenamefont {Hutson},\ and\ \citenamefont
		{Tarbutt}}]{Lim:PRA92:053419}%
	\BibitemOpen
	\bibfield  {author} {\bibinfo {author} {\bibfnamefont {J.}~\bibnamefont
			{Lim}}, \bibinfo {author} {\bibfnamefont {M.~D.}\ \bibnamefont {Frye}},
		\bibinfo {author} {\bibfnamefont {J.~M.}\ \bibnamefont {Hutson}},\ and\
		\bibinfo {author} {\bibfnamefont {M.~R.}\ \bibnamefont {Tarbutt}},\
	}\bibfield  {title} {\bibinfo {title} {Modeling sympathetic cooling of
			molecules by ultracold atoms},\ }\href
	{https://doi.org/10.1103/PhysRevA.92.053419} {\bibfield  {journal} {\bibinfo
			{journal} {Phys. Rev. A}\ }\textbf {\bibinfo {volume} {92}},\ \bibinfo
		{pages} {053419} (\bibinfo {year} {2015})},\ \Eprint
	{https://arxiv.org/abs/1508.03987} {arXiv:1508.03987 [physics]} \BibitemShut
	{NoStop}%
	\bibitem [{\citenamefont {Stuhl}\ \emph {et~al.}(2012)\citenamefont {Stuhl},
		\citenamefont {Hummon}, \citenamefont {Yeo}, \citenamefont
		{Qu{\'{e}}m{\'{e}}ner}, \citenamefont {Bohn},\ and\ \citenamefont
		{Ye}}]{Stuhl:Nature492:396}%
	\BibitemOpen
	\bibfield  {author} {\bibinfo {author} {\bibfnamefont {B.~K.}\ \bibnamefont
			{Stuhl}}, \bibinfo {author} {\bibfnamefont {M.~T.}\ \bibnamefont {Hummon}},
		\bibinfo {author} {\bibfnamefont {M.}~\bibnamefont {Yeo}}, \bibinfo {author}
		{\bibfnamefont {G.}~\bibnamefont {Qu{\'{e}}m{\'{e}}ner}}, \bibinfo {author}
		{\bibfnamefont {J.~L.}\ \bibnamefont {Bohn}},\ and\ \bibinfo {author}
		{\bibfnamefont {J.}~\bibnamefont {Ye}},\ }\bibfield  {title} {\bibinfo
		{title} {Evaporative cooling of the dipolar hydroxyl radical},\ }\href
	{https://doi.org/10.1038/nature11718} {\bibfield  {journal} {\bibinfo
			{journal} {Nature}\ }\textbf {\bibinfo {volume} {492}},\ \bibinfo {pages}
		{396} (\bibinfo {year} {2012})},\ \Eprint {https://arxiv.org/abs/1209.6343}
	{arXiv:1209.6343 [physics]} \BibitemShut {NoStop}%
	\bibitem [{\citenamefont {Ni}\ \emph {et~al.}(2008)\citenamefont {Ni},
		\citenamefont {Ospelkaus}, \citenamefont {de~Miranda}, \citenamefont {Pe'er},
		\citenamefont {Neyenhuis}, \citenamefont {Zirbel}, \citenamefont
		{Kotochigova}, \citenamefont {Julienne}, \citenamefont {Jin},\ and\
		\citenamefont {Ye}}]{Ni:Science322:231}%
	\BibitemOpen
	\bibfield  {author} {\bibinfo {author} {\bibfnamefont {K.-K.}\ \bibnamefont
			{Ni}}, \bibinfo {author} {\bibfnamefont {S.}~\bibnamefont {Ospelkaus}},
		\bibinfo {author} {\bibfnamefont {M.~H.~G.}\ \bibnamefont {de~Miranda}},
		\bibinfo {author} {\bibfnamefont {A.}~\bibnamefont {Pe'er}}, \bibinfo
		{author} {\bibfnamefont {B.}~\bibnamefont {Neyenhuis}}, \bibinfo {author}
		{\bibfnamefont {J.~J.}\ \bibnamefont {Zirbel}}, \bibinfo {author}
		{\bibfnamefont {S.}~\bibnamefont {Kotochigova}}, \bibinfo {author}
		{\bibfnamefont {P.~S.}\ \bibnamefont {Julienne}}, \bibinfo {author}
		{\bibfnamefont {D.}~\bibnamefont {Jin}},\ and\ \bibinfo {author}
		{\bibfnamefont {J.}~\bibnamefont {Ye}},\ }\bibfield  {title} {\bibinfo
		{title} {A high phase-space-density gas of polar molecules},\ }\href
	{https://doi.org/10.1126/science.1163861} {\bibfield  {journal} {\bibinfo
			{journal} {Science}\ }\textbf {\bibinfo {volume} {322}},\ \bibinfo {pages}
		{231} (\bibinfo {year} {2008})}\BibitemShut {NoStop}%
	\bibitem [{\citenamefont {P\'erez-R\'{\i}os}\ \emph {et~al.}(2015)\citenamefont
		{P\'erez-R\'{\i}os}, \citenamefont {Lepers},\ and\ \citenamefont
		{Dulieu}}]{Perez:PRL115:073201}%
	\BibitemOpen
	\bibfield  {author} {\bibinfo {author} {\bibfnamefont {J.}~\bibnamefont
			{P\'erez-R\'{\i}os}}, \bibinfo {author} {\bibfnamefont {M.}~\bibnamefont
			{Lepers}},\ and\ \bibinfo {author} {\bibfnamefont {O.}~\bibnamefont
			{Dulieu}},\ }\bibfield  {title} {\bibinfo {title} {Theory of long-range
			ultracold atom-molecule photoassociation},\ }\href
	{https://doi.org/10.1103/PhysRevLett.115.073201} {\bibfield  {journal}
		{\bibinfo  {journal} {Phys. Rev. Lett.}\ }\textbf {\bibinfo {volume} {115}},\
		\bibinfo {pages} {073201} (\bibinfo {year} {2015})},\ \Eprint
	{https://arxiv.org/abs/1505.03288} {arXiv:1505.03288 [physics]} \BibitemShut
	{NoStop}%
	\bibitem [{\citenamefont {Cheng}\ \emph {et~al.}(2016)\citenamefont {Cheng},
		\citenamefont {van~der Poel}, \citenamefont {Jansen}, \citenamefont
		{Quintero-P\'erez}, \citenamefont {Wall}, \citenamefont {Ubachs},\ and\
		\citenamefont {Bethlem}}]{Cheng:PRL117:253201}%
	\BibitemOpen
	\bibfield  {author} {\bibinfo {author} {\bibfnamefont {C.}~\bibnamefont
			{Cheng}}, \bibinfo {author} {\bibfnamefont {A.~P.~P.}\ \bibnamefont {van~der
				Poel}}, \bibinfo {author} {\bibfnamefont {P.}~\bibnamefont {Jansen}},
		\bibinfo {author} {\bibfnamefont {M.}~\bibnamefont {Quintero-P\'erez}},
		\bibinfo {author} {\bibfnamefont {T.~E.}\ \bibnamefont {Wall}}, \bibinfo
		{author} {\bibfnamefont {W.}~\bibnamefont {Ubachs}},\ and\ \bibinfo {author}
		{\bibfnamefont {H.~L.}\ \bibnamefont {Bethlem}},\ }\bibfield  {title}
	{\bibinfo {title} {Molecular fountain},\ }\href
	{https://doi.org/10.1103/PhysRevLett.117.253201} {\bibfield  {journal}
		{\bibinfo  {journal} {Phys. Rev. Lett.}\ }\textbf {\bibinfo {volume} {117}},\
		\bibinfo {pages} {253201} (\bibinfo {year} {2016})},\ \Eprint
	{https://arxiv.org/abs/1611.03640} {arXiv:1611.03640 [physics]} \BibitemShut
	{NoStop}%
	\bibitem [{Com()}]{Comsol:Multiphysics}%
	\BibitemOpen
	\href@noop {} {}\bibinfo {note} {COMSOL Multiphysics v.\ 5.5.\
		\url{http://www.comsol.com}. COMSOL AB, Stockholm, Sweden}\BibitemShut
	{NoStop}%
	\bibitem [{\citenamefont {Bethlem}\ \emph
		{et~al.}(2002{\natexlab{b}})\citenamefont {Bethlem}, \citenamefont
		{Crompvoets}, \citenamefont {Jongma}, \citenamefont {van~de Meerakker},\ and\
		\citenamefont {Meijer}}]{Bethlem:PRA65:053416}%
	\BibitemOpen
	\bibfield  {author} {\bibinfo {author} {\bibfnamefont {H.~L.}\ \bibnamefont
			{Bethlem}}, \bibinfo {author} {\bibfnamefont {F.~M.~H.}\ \bibnamefont
			{Crompvoets}}, \bibinfo {author} {\bibfnamefont {R.~T.}\ \bibnamefont
			{Jongma}}, \bibinfo {author} {\bibfnamefont {S.~Y.~T.}\ \bibnamefont {van~de
				Meerakker}},\ and\ \bibinfo {author} {\bibfnamefont {G.}~\bibnamefont
			{Meijer}},\ }\bibfield  {title} {\bibinfo {title} {Deceleration and trapping
			of ammonia using time-varying electric fields},\ }\href
	{https://doi.org/10.1103/PhysRevA.65.053416} {\bibfield  {journal} {\bibinfo
			{journal} {Phys. Rev. A}\ }\textbf {\bibinfo {volume} {65}},\ \bibinfo
		{pages} {053416} (\bibinfo {year} {2002}{\natexlab{b}})}\BibitemShut
	{NoStop}%
	\bibitem [{\citenamefont {Yurchenko}\ \emph {et~al.}(2011)\citenamefont
		{Yurchenko}, \citenamefont {Barber},\ and\ \citenamefont
		{Tennyson}}]{Yurchenko:MNRAS413:1828}%
	\BibitemOpen
	\bibfield  {author} {\bibinfo {author} {\bibfnamefont {S.~N.}\ \bibnamefont
			{Yurchenko}}, \bibinfo {author} {\bibfnamefont {R.~J.}\ \bibnamefont
			{Barber}},\ and\ \bibinfo {author} {\bibfnamefont {J.}~\bibnamefont
			{Tennyson}},\ }\bibfield  {title} {\bibinfo {title} {A variationally computed
			line list for hot {NH}$_3$},\ }\href
	{https://doi.org/10.1111/j.1365-2966.2011.18261.x} {\bibfield  {journal}
		{\bibinfo  {journal} {Mon. Not. R. Astron. Soc.}\ }\textbf {\bibinfo {volume}
			{413}},\ \bibinfo {pages} {1828} (\bibinfo {year} {2011})},\ \Eprint
	{https://arxiv.org/abs/1011.1569} {arXiv:1011.1569 [astro-ph]} \BibitemShut
	{NoStop}%
	\bibitem [{\citenamefont {Yurchenko}\ \emph {et~al.}(2009)\citenamefont
		{Yurchenko}, \citenamefont {Barber}, \citenamefont {Yachmenev}, \citenamefont
		{Thiel}, \citenamefont {Jensen},\ and\ \citenamefont
		{Tennyson}}]{Yurchenko:JPCA113:11845}%
	\BibitemOpen
	\bibfield  {author} {\bibinfo {author} {\bibfnamefont {S.~N.}\ \bibnamefont
			{Yurchenko}}, \bibinfo {author} {\bibfnamefont {R.~J.}\ \bibnamefont
			{Barber}}, \bibinfo {author} {\bibfnamefont {A.}~\bibnamefont {Yachmenev}},
		\bibinfo {author} {\bibfnamefont {W.}~\bibnamefont {Thiel}}, \bibinfo
		{author} {\bibfnamefont {P.}~\bibnamefont {Jensen}},\ and\ \bibinfo {author}
		{\bibfnamefont {J.}~\bibnamefont {Tennyson}},\ }\bibfield  {title} {\bibinfo
		{title} {A variationally computed {$T=300$~K} line list for {NH}$_3$},\
	}\href {https://doi.org/10.1021/jp9029425} {\bibfield  {journal} {\bibinfo
			{journal} {J. Phys. Chem. A}\ }\textbf {\bibinfo {volume} {113}},\ \bibinfo
		{pages} {11845} (\bibinfo {year} {2009})}\BibitemShut {NoStop}%
	\bibitem [{\citenamefont {Yurchenko}\ \emph {et~al.}(2007)\citenamefont
		{Yurchenko}, \citenamefont {Thiel},\ and\ \citenamefont
		{Jensen}}]{Yurchenko:JMS245:126}%
	\BibitemOpen
	\bibfield  {author} {\bibinfo {author} {\bibfnamefont {S.~N.}\ \bibnamefont
			{Yurchenko}}, \bibinfo {author} {\bibfnamefont {W.}~\bibnamefont {Thiel}},\
		and\ \bibinfo {author} {\bibfnamefont {P.}~\bibnamefont {Jensen}},\
	}\bibfield  {title} {\bibinfo {title} {Theoretical {ROVibrational} energies
			({TROVE}): A robust numerical approach to the calculation of rovibrational
			energies for polyatomic molecules},\ }\href
	{https://doi.org/10.1016/j.jms.2007.07.009} {\bibfield  {journal} {\bibinfo
			{journal} {J. Mol. Spectrosc.}\ }\textbf {\bibinfo {volume} {245}},\ \bibinfo
		{pages} {126} (\bibinfo {year} {2007})}\BibitemShut {NoStop}%
	\bibitem [{\citenamefont {Yachmenev}\ and\ \citenamefont
		{Yurchenko}(2015)}]{Yachmenev:JCP143:014105}%
	\BibitemOpen
	\bibfield  {author} {\bibinfo {author} {\bibfnamefont {A.}~\bibnamefont
			{Yachmenev}}\ and\ \bibinfo {author} {\bibfnamefont {S.~N.}\ \bibnamefont
			{Yurchenko}},\ }\bibfield  {title} {\bibinfo {title} {Automatic
			differentiation method for numerical construction of the
			rotational-vibrational {H}amiltonian as a power series in the curvilinear
			internal coordinates using the {E}ckart frame},\ }\href
	{https://doi.org/10.1063/1.4923039} {\bibfield  {journal} {\bibinfo
			{journal} {J. Chem. Phys.}\ }\textbf {\bibinfo {volume} {143}},\ \bibinfo
		{pages} {014105} (\bibinfo {year} {2015})}\BibitemShut {NoStop}%
	\bibitem [{\citenamefont {Yurchenko}\ \emph {et~al.}(2017)\citenamefont
		{Yurchenko}, \citenamefont {Yachmenev},\ and\ \citenamefont
		{Ovsyannikov}}]{Yurchenko:JCTC13:4368}%
	\BibitemOpen
	\bibfield  {author} {\bibinfo {author} {\bibfnamefont {S.~N.}\ \bibnamefont
			{Yurchenko}}, \bibinfo {author} {\bibfnamefont {A.}~\bibnamefont
			{Yachmenev}},\ and\ \bibinfo {author} {\bibfnamefont {R.~I.}\ \bibnamefont
			{Ovsyannikov}},\ }\bibfield  {title} {\bibinfo {title} {Symmetry adapted
			ro-vibrational basis functions for variational nuclear motion calculations:
			{TROVE} approach},\ }\href {https://doi.org/10.1021/acs.jctc.7b00506}
	{\bibfield  {journal} {\bibinfo  {journal} {J. Chem. Theory\ Comput.}\
		}\textbf {\bibinfo {volume} {13}},\ \bibinfo {pages} {4368} (\bibinfo {year}
		{2017})},\ \Eprint {https://arxiv.org/abs/1708.07185} {arXiv:1708.07185
		[physics]} \BibitemShut {NoStop}%
	\bibitem [{\citenamefont {Owens}\ and\ \citenamefont
		{Yachmenev}(2018)}]{Owens:JCP148:124102}%
	\BibitemOpen
	\bibfield  {author} {\bibinfo {author} {\bibfnamefont {A.}~\bibnamefont
			{Owens}}\ and\ \bibinfo {author} {\bibfnamefont {A.}~\bibnamefont
			{Yachmenev}},\ }\bibfield  {title} {\bibinfo {title} {{RichMol}: A general
			variational approach for rovibrational molecular dynamics in external
			electric fields},\ }\href {https://doi.org/10.1063/1.5023874} {\bibfield
		{journal} {\bibinfo  {journal} {J. Chem. Phys.}\ }\textbf {\bibinfo {volume}
			{148}},\ \bibinfo {pages} {124102} (\bibinfo {year} {2018})},\ \Eprint
	{https://arxiv.org/abs/1802.07603} {arXiv:1802.07603 [physics]} \BibitemShut
	{NoStop}%
	\bibitem [{\citenamefont {Scoles}(1988)}]{Scoles:MolBeam:1}%
	\BibitemOpen
	\bibinfo {editor} {\bibfnamefont {G.}~\bibnamefont {Scoles}},\ ed.,\
	\href@noop {} {\emph {\bibinfo {title} {Atomic and molecular beam
				methods}}},\ Vol.~\bibinfo {volume} {1}\ (\bibinfo  {publisher} {Oxford
		University Press},\ \bibinfo {address} {New York, NY, USA},\ \bibinfo {year}
	{1988})\BibitemShut {NoStop}%
	\bibitem [{\citenamefont {Christen}\ \emph {et~al.}(2010)\citenamefont
		{Christen}, \citenamefont {Rademann},\ and\ \citenamefont
		{Even}}]{Christen:JPCA114:11189}%
	\BibitemOpen
	\bibfield  {author} {\bibinfo {author} {\bibfnamefont {W.}~\bibnamefont
			{Christen}}, \bibinfo {author} {\bibfnamefont {K.}~\bibnamefont {Rademann}},\
		and\ \bibinfo {author} {\bibfnamefont {U.}~\bibnamefont {Even}},\ }\bibfield
	{title} {\bibinfo {title} {Supersonic beams at high particle densities: Model
			description beyond the ideal gas approximation},\ }\href
	{https://doi.org/10.1021/jp102855m} {\bibfield  {journal} {\bibinfo
			{journal} {J. Phys. Chem. A}\ }\textbf {\bibinfo {volume} {114}},\ \bibinfo
		{pages} {11189} (\bibinfo {year} {2010})}\BibitemShut {NoStop}%
	\bibitem [{\citenamefont {Schnell}\ and\ \citenamefont
		{Meijer}(2009)}]{Schnell:ACIE48:6010}%
	\BibitemOpen
	\bibfield  {author} {\bibinfo {author} {\bibfnamefont {M.}~\bibnamefont
			{Schnell}}\ and\ \bibinfo {author} {\bibfnamefont {G.}~\bibnamefont
			{Meijer}},\ }\bibfield  {title} {\bibinfo {title} {Cold molecules:
			Preparation, applications, and challenges},\ }\href
	{https://doi.org/10.1002/anie.200805503} {\bibfield  {journal} {\bibinfo
			{journal} {Angew. Chem. Int. Ed.}\ }\textbf {\bibinfo {volume} {48}},\
		\bibinfo {pages} {6010} (\bibinfo {year} {2009})}\BibitemShut {NoStop}%
	\bibitem [{\citenamefont {Yachmenev}\ and\ \citenamefont
		{K\"{u}pper}(2017)}]{Yachmenev:JCP147:141101}%
	\BibitemOpen
	\bibfield  {author} {\bibinfo {author} {\bibfnamefont {A.}~\bibnamefont
			{Yachmenev}}\ and\ \bibinfo {author} {\bibfnamefont {J.}~\bibnamefont
			{K\"{u}pper}},\ }\bibfield  {title} {\bibinfo {title} {Communication: General
			variational approach to nuclear-quadrupole coupling in rovibrational spectra
			of polyatomic molecules},\ }\href {https://doi.org/10.1063/1.5002533}
	{\bibfield  {journal} {\bibinfo  {journal} {J. Chem. Phys.}\ }\textbf
		{\bibinfo {volume} {147}},\ \bibinfo {pages} {141101} (\bibinfo {year}
		{2017})},\ \Eprint {https://arxiv.org/abs/1709.08558} {arXiv:1709.08558
		[physics]} \BibitemShut {NoStop}%
	\bibitem [{\citenamefont {Zhang}\ \emph {et~al.}(2016)\citenamefont {Zhang},
		\citenamefont {Meijer},\ and\ \citenamefont
		{Vanhaecke}}]{Zhang:PRA93:023408}%
	\BibitemOpen
	\bibfield  {author} {\bibinfo {author} {\bibfnamefont {D.}~\bibnamefont
			{Zhang}}, \bibinfo {author} {\bibfnamefont {G.}~\bibnamefont {Meijer}},\ and\
		\bibinfo {author} {\bibfnamefont {N.}~\bibnamefont {Vanhaecke}},\ }\bibfield
	{title} {\bibinfo {title} {Advanced switching schemes in a {S}tark
			decelerator},\ }\href {https://doi.org/10.1103/PhysRevA.93.023408} {\bibfield
		{journal} {\bibinfo  {journal} {Phys. Rev. A}\ }\textbf {\bibinfo {volume}
			{93}},\ \bibinfo {pages} {023408} (\bibinfo {year} {2016})},\ \Eprint
	{https://arxiv.org/abs/1512.08361} {arXiv:1512.08361 [physics]} \BibitemShut
	{NoStop}%
\end{thebibliography}
\end{document}